\documentclass[a4paper]{article}
\usepackage{amsmath}
\usepackage{mathtools}
\usepackage{amssymb}
\usepackage{latexsym}
\usepackage[utf8]{inputenc}
\usepackage{xcolor}
\usepackage{natbib}
\usepackage{graphicx}
\usepackage{float}
\usepackage{caption}
\usepackage{subcaption}
\usepackage{soul}
\usepackage{amsthm}
\newcommand\itN {{\mathcal{N}}}

\def\natu{\mathbb{N}}

\newcommand{\esp}{\mathbb{E}}
\newcommand{\prob}{\mathbb{P}}

	\newcommand{\convprob  }{ \buildrel{p}\over\longrightarrow}

\newcommand{\convdist}{ \buildrel{D}\over\longrightarrow}

\def\square{\ifmmode\sqr\else{$\sqr$}\fi}
\def\sqr{\vcenter{
         \hrule height.1mm
         \hbox{\vrule width.1mm height2.2mm\kern2.18mm
\vrule width.1mm}
         \hrule height.1mm}}

\theoremstyle{plain}
{\newcommand{\url}{\text}}

\newtheorem*{theorem*}{Theorem}
\newtheorem{theorem}{Theorem}

\title{An unbiased estimator of the case fatality rate}
\author{Agustín Alvarez$^{(1)}$, Marina Fragalá$^{(1, 2)}$ and Marina Valdora$^{(2, 3)}$}

\date{}

\begin{document}
	\maketitle
	{\small (1)  Instituto de Ciencias, Universidad de General Sarmiento}
	
	{\small (2) Departamento de matemática, Facultad de Ciencias Exactas y Naturales, Universidad de Buenos Aires}	
	
	{\small (3) Instituto de Cálculo, Universidad de Buenos Aires - CONICET}	
	
	{\small Corresponding author: 
		
		E-mail: agalvarez@campus.ungs.edu.ar
		
		Tel.: +54-11-44697500 (7207)
		
		Address: Juan María Gutierrez 1150, Los Polvorines, Buenos Aires, Argentina.
		
		ORCID: 0000-0003-4359-4059
	
}
	
		{\small }
	
	\begin{abstract}
During an epidemic outbreak of a new disease, the probability of dying once infected is considered an important though difficult task to be computed. Since it is very hard to know the true number of infected people, the focus is placed on estimating the case fatality rate, which is defined as the probability of dying once tested and confirmed as infected. The estimation of this rate at the beginning of an epidemic remains challenging for several reasons, including the time gap between diagnosis and death, and the rapid growth in the number of confirmed cases.

In this work, an unbiased estimator of the case fatality rate of a virus is presented. The consistency of the estimator is demonstrated, and its asymptotic distribution is derived, enabling the corresponding confidence intervals (C.I.) to be established. The proposed method is based on the distribution $F$ of the time between confirmation and death of individuals who die because of the virus. The estimator's performance is analyzed in both simulation scenarios and the real-world context of Argentina in $2020$ for the COVID-19 pandemic, consistently achieving excellent results when compared to an existing proposal 
 as well as to the conventional ``naive'' estimator that was employed to report the case fatality rates during the last COVID-19 pandemic.

In the simulated scenarios, the empirical coverage of our C.I. is studied, both using the $F$ employed to generate the data and an estimated $F$, and it is observed that the desired level of confidence is reached quickly when using real $F$ and in a reasonable period of time when estimating $F$.
\end{abstract}
	\section{Introduction}

One of the most important questions to be answered when a new infectious disease emerges, such as COVID-19, is how deadly it is or will be. In other words, the proportion of infected individuals who will die as the epidemic progresses needs to be determined. This proportion is considered a key epidemiological measure for quantifying the severity of the disease, and it is particularly crucial to have it estimated during outbreaks of emerging infectious diseases like COVID-19.

To calculate this rate, the number of infected individuals needs to be known, which is not a trivial matter. Typically, the data consist of individuals who were tested and confirmed positive for the infection in question, referred to as "confirmed cases" from now on.

In massive infections like COVID-19, the actual number of infected individuals is often unknown due to the presence of asymptomatic cases, cases that were not tested, and a lack of serological studies, among other reasons. Given this issue, the fatality rate among confirmed cases is commonly studied, as done in this article. Other authors who follow the same criterion are, e.g., \cite{marschner2021case} and \cite{grewelle2020estimating}.

The proportion of confirmed cases that die from the disease during an epidemic is difficult to calculate for several reasons: rapid growth in the number of confirmed cases, the time gap between diagnosis and death, biases due to delays in reporting confirmed cases, among others (see \cite{baud2020real}).

In epidemiology, the case fatality rate among confirmed cases in a specific period of time (a day, a week, a month, etc.) is defined as the proportion of confirmed individuals in that period who eventually die (not necessarily within that period) due to the disease. In this work, a mathematical definition of the case fatality rate among confirmed cases for a specific day is provided as the probability to die from the disease for a randomly chosen individual among the confirmed cases from the beginning of the epidemic up to that day. This means that the time periods considered are of the form $[0, t]$, where $t\in \mathbb{N}_0$, with 0 representing the day when the first case in the geographic region of interest is recorded. We will refer to that day as day 0 of the epidemic. The case fatality rate for the period $[0,t]$ is denoted $cfr(t)$ and is the object of estimation.

A commonly calculated ``naive'' fatality rate is the proportion of confirmed individuals during a fixed period of time who die from the disease in that same period. The World Health Organization and many countries reported this ``naive'' fatality rate daily for COVID-19, considering the period of time from the beginning of the epidemic up to the reporting day. The reason for using this rate is that it requires minimal information for calculation (see Kim et al. 2021). This rate has a tendency to underestimate $cfr(t)$ because, up to the reporting date, many of the confirmed cases have not died yet. This fact has been described by several authors, who also have made attempts to define and estimate $cfr(t)$; see, for example, Chang et al. (2020) and Shim et al. (2020). Lipsich et al. (2015) and Marschner (2021) also define the case fatality rate as the probability to die from the disease for a randomly chosen individual confirmed within a fixed period of time. These authors analyze potential biases in their estimates. Lee and Lim (2019) also make an attempt in that direction.

The ``naive'' fatality rate, reported daily during the COVID-19 pandemic, underestimates $cfr(t)$ because the calculation for a specific day involves dividing the number of people who died from COVID-19 until that day by the number of confirmed cases until that day. The underestimation is because the numerator does not include the confirmed cases that will die from the disease later on. This bias can be significant, especially when the estimation is made during a period of rapid growth in confirmed cases or when the time between diagnosis and death is long. For example, at the beginning of an outbreak, the number of confirmed cases can double in just a few days, but only a small proportion of the patients who will eventually die do so in the first few days after diagnosis.

To address this underestimation issue of the ``naive'' rate,  Garske et al. (2009) propose an estimator that takes into account the distribution of the time between confirmation and death for individuals who die from the disease. The method they propose would be unbiased if the daily probabilities of a confirmed individual dying from the disease did not change over time. However, this probability can change from one day to another for various reasons, not only because a treatment that reduces the probability is found but also, for example, because the definition of a confirmed case changes or because more testing becomes accessible. These last two reasons, which were very common during the COVID-19 pandemic, do not change the fatality rate among infected individuals but do change fatality rate among confirmed individuals. The fact that the daily probabilities of dying among confirmed cases vary makes Garke's estimator biased.

In this work an estimator is proposed that remains unbiased for $cfr(t)$ even when the daily probabilities of death among confirmed cases are not constant over time. Both the \cite{garske2009assessing} estimator and ours assume that the distribution of the time between confirmation and death for individuals who die from the disease is known.

Our proposal, the Garske et al. estimator, and the ``naive'' estimator are applied and compared with real COVID-19 data in Argentina during 2020. Different simulation scenarios are also considered. A very good performance of our proposal compared to the other estimation methods is observed, both for real data and in simulations. The consistency of our estimator is proven, and its asymptotic distribution is found, allowing for the derivation of confidence intervals for $cfr(t)$. In the simulation scenarios, the level of empirical coverage of confidence intervals is studied when our estimator does not assume the distribution of the time between confirmation and death is known but estimates it. The finite sample bias and the mean squared error of our estimator, \cite{garske2009assessing}'s estimator, and the ``naive'' rate are analyzed, observing a better performance of our estimator in all cases.

The real data analysis is based on a data base that was published and updated daily by the Ministry of Health of Argentina since March 2020 until at least the end of 2021. It contains information on all the individuals that were tested for COVID-19 during that time. For each individual, it provides sex, age, country of residence, province, date of first symptoms, date of diagnosis and date of death, among other features.


\section{Proposed estimators}

\subsection{Main definitions and notation}\label{sec:defi}

Random variables will be denoted with upper case letters, and non-random parameters will be denoted with lower case letters. We set the following definitions:

\begin{itemize}
	\item $ p_d$ is the probability of death among cases confirmed during day $d$.
	\item $c_d$ is the number of cases confirmed during day $d$.
	\item $D_{d,i}$ is a dichotomous variable that equals $1$ if the $i$--th confirmed case on day $d$ dies {because of the disease} and $0$ if it does not.
	\item $D_{d,i}(t)$ is a dichotomous variable that equals $1$ if the $i$--th confirmed case on day $d$ has died {because of the disease} by day $t$ and $0$ if it has not, defined for $d\leq t$.
	\item $D_{e} (t)$ is the total number of people that die from COVID-19 infection among cases confirmed until day $t$ included, once the epidemic has ended.
	\item {$D(t)$ is the number of confirmed cases that died from COVID-19 from the beginning of the epidemic (day $0$) until day $t$ inclusive.}
\end{itemize}
Suppose that $D_{d,1},\ldots,D_{d,c_d}$ are independent random variables with a Bernoulli distribution and probability of success $p_d$, \textsl{i.e.} $D_{d,i}\sim {Be}(p_d)$. Thus,

\begin{equation} \label{eq:mftymt}
	D_{e}(t)=\sum\limits_{d=0}^t\sum\limits_{i=1}^{c_d}D_{d,i} \,\, \text{ and } \,\, D(t)=\sum_{d=0}^{t} \sum_{i=1}^{c_d} D_{d, i}(t).
\end{equation}

The ``naive'' estimator of the case fatality rate usually reported on day $t$, denoted as $CFR_N(t)$, is defined as:

\begin{eqnarray}\label{eq:naive}
	CFR_N(t)=\frac{D(t)}{\displaystyle\sum_{d=0}^t c_d}.
\end{eqnarray}

$CFR_F(t)$ is defined as the proportion of cases confirmed until day $t$ which finally die because of the disease, and referred to as the final case fatality rate by day $t$. In terms of the defined variables:
\begin{eqnarray}\label{final}
	CFR_F(t)=\frac{D_e(t)}{\displaystyle\sum_{d=0}^t c_d}.
\end{eqnarray}
Notice that $CFR_F(t)$ cannot be computed on day $t$; one would have to wait for all the diagnosed people by day $t$ to recover or die.
We define the {case fatality rate} by day $t$ as the expected value of $CFR_F(t)$, \textsl{i.e},
$cfr(t)=\esp\left(CFR_F(t)\right)$. As it will be seen in \eqref{eq:weightedsum}, this definition of $cfr(t)$ coincides with the one given in the Introduction. 
It is worth noting that the {case fatality rate} by $t$ defined in this way is a population {parameter} and it is not observable, not even at the end of the epidemic. From \eqref{eq:mftymt} $$\esp(D_e(t)) = \sum\limits_{d=0}^t c_d p_d,$$ and then, from \eqref{final}
\begin{equation}\label{eq:weightedsum}
	cfr(t)=\frac{\displaystyle\sum\limits_{d=0}^t c_d p_d}{\displaystyle\sum\limits_{d=0}^t c_d}=\sum\limits_{d=0}^t \omega_d p_d,
\end{equation} where $$w_d= \frac {c_d}{\displaystyle\sum_{d=0}^t c_d}.$$
 Note that $cfr(t)$ is a weighted sum of daily case fatality rates $p_d$, where each day's weight is the proportion of cases that were confirmed that day with respect to the total number of cases confirmed until day $t$. Thus, $cfr(t)$ can be interpreted as the probability of dying from the disease for a randomly picked person among those confirmed by day $t$. If $p_d=p$ (constant throughout the epidemic), $cfr(t)=p$. It is worth noting that $cfr(t)$ is the parameter of interest, of which the case fatality rate observed at the end of the epidemic, $CFR_F(t)$, is an estimation.

In order to estimate $cfr(t)$, the following definition is made:

\noindent $T_{d,i}:=$ number of days since confirmation until death, in the $i$-th case confirmed on day $d$.

Let $F_d$ be the cumulative distribution function of $T_{d,i}$ conditional on \linebreak $D_{d,i}=1$. 
\noindent Several authors use this distribution in their estimations, see for instance \cite{marschner2021case}, \cite{garske2009assessing}, \cite{nishiura2009early}, \cite{dorigatti2020report}. Note that we are allowing the cumulative distribution function to change in time, unlike \cite{garske2009assessing}. This seems realistic for several reasons; for instance, new treatments may lengthen the survival time, the sanitary system may collapse and this may shorten the survival time, or the confirmation of cases may be faster, lengthening also the time from confirmation until death, among others.


\subsection{An unbiased estimator for $\boldsymbol{cfr(t)}$}\label{sec:estimators}

Our goal is to {predict} on day $t$ the number of people that eventually will die, among confirmed cases until day $t$, that is to say $D_e(t)$, as defined in \eqref{eq:mftymt}. 
The idea of our proposal is the following: we know that, among the cases confirmed on day $d$, with $d\le t$, the expected proportion that will have died by day $t$, among those  that have died or will die eventually is $F_d(t-d)$. Therefore, dividing 
the cases confirmed during day $d$ that have  died by day $t$ by $F_d(t-d)$, we will obtain a {predictor} of the number of confirmed cases during day $d$ that will finally die. In concrete, we define the predictor of $D_e(t)$ as follows

$$\widehat{D}_e(t)=\displaystyle\sum_{d=0}^t\displaystyle\frac{ \sum\limits_{i=1}^{c_d}D_{d,i}(t)}{F_d(t-d)}.$$
The estimator of the case fatality rate by day $t$, that is the estimator of $cfr(t)$, is defined as
\begin{equation}\label{eq:cfrnuestra}
CFR(t)=\frac{\widehat{D}_e(t)}{\displaystyle\sum_{d=0}^t c_d}.
\end{equation}
Note that the probability of dying from COVID-19 by day $t$ for a case confirmed on day $d$ can be expressed in the following way
\begin{eqnarray}\nonumber
	\prob(D_{d,i}(t)=1)&=& \prob(D_{d,i}=1)\cdot \prob(T_{d,i}\le t-d|D_{d,i}=1)\\\nonumber	&=&p_dF_d(t-d).
\end{eqnarray}
Straightforward calculations show that $\esp(CFR(t))= cfr(t)$ and therefore, the proposed estimator is unbiased. Setting $$Z_{d,i}(t)=\frac{D_{d,i}(t)}{F_d(t-d)},$$ $CFR(t)$ is an average of the $Z_{d,i}(t)$ variables, which are independent but not necessarily identically distributed since  $D_{d,i}(t)\sim Be(p_d F_d(t-d))$. 


The following result states the consistency and asymptotic normality of $CFR(t)$. Its proof can be found in Appendix \ref{apendice}. Consider the following assumptions:
$$\begin{array}{llll}
	\textbf{A1}: & D &  \coloneqq & \displaystyle\inf_{d} F_{d}(0) > 0 \\
    \textbf{A2}: & I &  \coloneqq & \displaystyle\inf_{d} p_{d} > 0\\ 
    \textbf{A3}: & S &  \coloneqq & \displaystyle\sup_{d} p_{d} <1
\end{array}$$ 
\begin{theorem} \label{teo}
	For each $t\in\natu$, let  $\{D_{d,i}(t)\}_{d,i}$ for $0\le d\le t $ and $1\le i\le c_d$ be independent random variables ${Be(p_d F_d(t-d))}$. Assume that the total number of confirmed cases until day $t$, $$r_t\coloneqq\sum\limits_{d=0}^t c_d\stackrel{t\to\infty}{\longrightarrow} \infty.$$ Then
	
	\noindent $(i)$ If \textbf{A1} holds, then $CFR(t)-cfr(t) \convprob  0\,,$
	
	\noindent $(ii)$ If \textbf{A1} to \textbf{A3} hold, then  $\displaystyle\frac{CFR(t)-cfr(t)}{\sqrt{\mathbb{V}(CFR(t))}}\stackrel{D}{\longrightarrow} N(0,1),$ \\ 
	where $\mathbb{V}(CFR(t)$ is the variance of the $CFR(t)$.
\end{theorem} 
As a consequence, can be developed an asymptotic $(1-\alpha) \cdot 100 \%$ confidence interval for the expected value of the average, $cfr(t)$. Confidence bounds can be calculated by using a normal
approximation $$CFR(t)\pm z_{1-\frac{\alpha} {2}}\sqrt{\mathbb{V}(CFR(t))},$$ where 
\begin{equation}
\mathbb{V}(CFR(t))=\frac {1}{ r_t^2}\displaystyle\sum_{d=0}^t \frac{c_d p_d(1- p_dF_d(t-d))}{F_d(t-d)}
\end{equation}

Note that these confidence intervals are theoretical, in the sense that they only can be computed if $F_d$ and $p_d$ are known. 
Also the estimator $CFR(t)$ can be computed if $F_d$ is known. Since this is not the case when working with real data, in that case it is replaced by an estimator $\hat F_d$, see Section \ref{sec:F}. 
Therefore, the predictor that will be used for real data is given by
$$\widehat{D}_e(t)=\sum\limits_{d=0}^t\frac{ \sum\limits_{i=1}^{c_d}D_{d,i}(t)}{{\hat F_d(t-d)}}.$$
To calculate the confidence interval on day $t$, the $ p_d$ values are also needed for $0\le d\le t$, which are not available in a real data analysis. In such case, for $t_0\le t_1$, call $cfr_{t_0}^{t_1}$  the expected proportion of people that would die from COVID-19 between those who are confirmed in the period of time from $t_0$ to $t_1$ days after the start of the epidemic. In order to estimate $p_{d'}$ on day $t$, for $d'\le t $ is proposed an estimator of $cfr_{d'-3}^{d'+3}$ based on the same idea of the cumulative case fatality rate estimator $CFR$, as follows
\begin{eqnarray}\label{semanal}
	\hat{p}_{d'}= CFR_{d'-3}^{d'+3}(t)=\displaystyle\frac{\displaystyle\sum_{d=d'-3}^{d'+3}\frac{\sum\limits_{i=1}^{c_d} D_{d,i}(t)}{F_d(t-d)}}{\displaystyle\sum_{d=d'-3}^{d'+3} c_d}\,.
\end{eqnarray}
Estimation \eqref{semanal} is computed for days $d'$, with $3\le d'\le t-3 $, while we set $\hat{p}_{d'}=\hat{p}_{t-3}$ for $t-2\le d'\le t$ and 
$\hat{p}_{d'}=\hat{p}_{3}$ for $0\le d'\le 2$. Here, the estimation of $p_d$ is used  as an auxiliary calculus in order to estimate the variance of the confidence intervals. It is a problem of interest in its own since it allows to estimate the actual probability of dying from COVID-19 for the cases confirmed on day $d$. An analogous calculation can be done in order to estimate the daily probability of needing an intensive care unit (ICU), which might be useful to predict the number of people who will arrive to ICU. {This may be the subject of future work.}
%


\subsection{Modified \cite{garske2009assessing} estimator}

\cite{garske2009assessing} do not consider the possibility that the distribution of the time from confirmation to death, $F_d$, or the daily case fatality rate $ p_d$, may vary with time. Their estimator of the case fatality rate is defined as follows
\begin{eqnarray}\label{eq:garske}
	CFR_G(t)=\frac{D(t)}{\displaystyle\sum_{d=0}^t c_dF(t-d)},
\end{eqnarray} 
where $F$ is the distribution of the time from confirmation to death and $D(t)$  is the random variable that counts the number of confirmed cases that died from the beginning of the epidemic (day $0$), to day $t$. If we consider varying $F_d$ their estimator becomes
\begin{eqnarray}\nonumber
CFR_G^D(t)=\frac{D(t)}{\displaystyle\sum_{d=0}^t c_dF_d(t-d)}.
\end{eqnarray} 
Simple computations yield
\begin{eqnarray}\label{garskeEsp}
\esp(CFR_G^D(t))=\frac{\displaystyle\sum_{d=0}^t c_dF_d(t-d) p_d}{\displaystyle\sum_{d=0}^t c_dF_d(t-d)}=\sum\limits_{d=0}^t \omega^*_d p_d.
\end{eqnarray}
Equation \eqref{garskeEsp} shows that the expected value of this estimator is a weighted mean of $p_d$, but the weights are not the same as in $ cfr(t)$ and therefore, the estimator is not necessarily unbiased, unless $p_d=p$ for all $d$. Of course, these estimators are also computed by replacing $F$ or $F_d$ by their estimates when these distributions are not known.

{Despite this bias, Garske et al. estimator has some advantages if $p_d$ is constant. First, if the distribution function  $F_d = F$ is known or estimated previously, it can be computed for day $t$ using only the number of confirmed cases and the number of deaths until day $t$. The estimator proposed in our work, on the other hand, requires knowing the number of deaths by day $t$, among all confirmed cases during day $d$, for all $d\leq t$.}


\subsection{Estimation of $\boldsymbol{F_d}$}\label{sec:F}
Briefly, an explanation is provided on how $F_d$ is estimated on day $t$, for $d\leq t$. For such estimation, the assumption made is that $F_d\equiv F$ for all $d$. On day $t$, $F_d(k)$ is estimated  by $\hat F_{\textsc{emp}}$, the proportion of confirmed cases who died in $k$ or less days since confirmation, among those who died form the disease. For calculating this proportion we consider only cases who were confirmed by day $t-t_{back}$ and dead by day $t$, where $t_{back}$ is taken in such a way that the probability that a confirmed case that finally dies, does it in $t_{back}$ days or less, is high. The need to take this $t_{back}$ value is that considering all cases confirmed until day $t$ would lead to an overestimation of $F(k)$, for small values of $k$, since if there are dead people for day $t$, among the cases confirmed in the last days before $t$, they have inevitably died within a few days, while those who have been infected a few days before $t$ and have not died by day $t$ but will delay some more days to die, are not taken into account for the estimate of $F$. Ideally, $t_{back}$ should be chosen in such a way that the probability of dying in $t_{back}$ days or less, among people who die from the disease,  is one. 
If one chooses to set the value of $t_{back}$ to a very large number, the estimate of $F$ would have a very small bias. However, one would need to wait for many days to pass before being able to start estimating $F$ and consequently $cfr(t)$.
Let us see an example of how the value of $t_{back}$ can be determined in practice:
suppose we are on day 75 of the epidemic and we analyze those people who were confirmed infected in the first 30 days of the epidemic, in other words, we use $t_{back}=45$. If we observe that $95\%$ of them have already recovered, and of the remaining $5\%$, $98\%$ have already died from the disease, we would be able to estimate that the probability of dying from the disease in 45 days or less, among those who die from the disease, is at least $98\%$. 
In the choice of $t_{back}$, there is a trade-off between achieving a low-bias estimation of $F$ and being able to do it as early as possible. In our case, we have chosen a lower bound of $98\%$ for the probability of dying in $t_{back}$ days or less, conditional on dying from the disease.



		\section{Monte Carlo study}\label{sec:montecarlo}
\begin{figure}[H]
	\centering
	\begin{subfigure}{0.5\textwidth}
		\centering
		\includegraphics[width=0.9\linewidth]{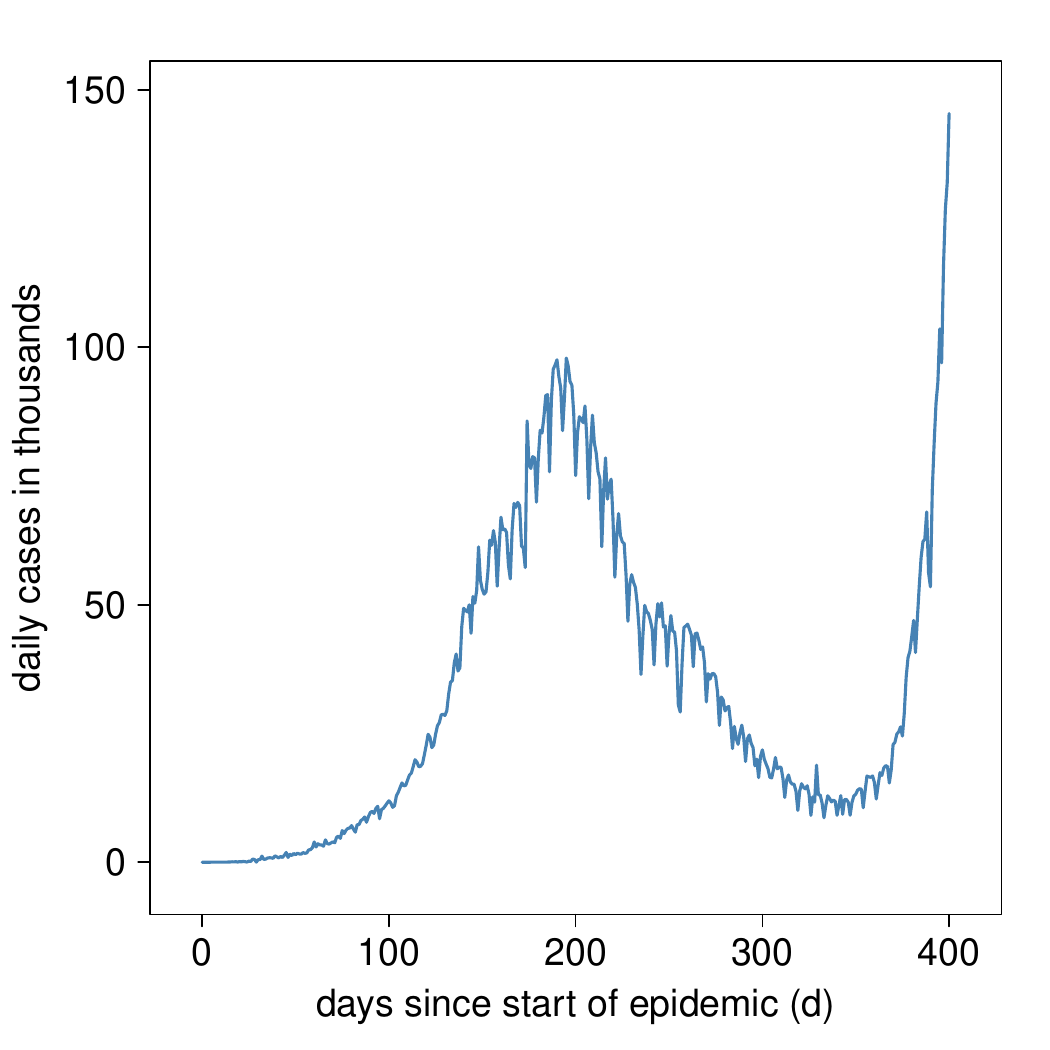}
	\end{subfigure}%
	\begin{subfigure}{0.5\textwidth}
		\centering
		\includegraphics[width=0.9\linewidth]{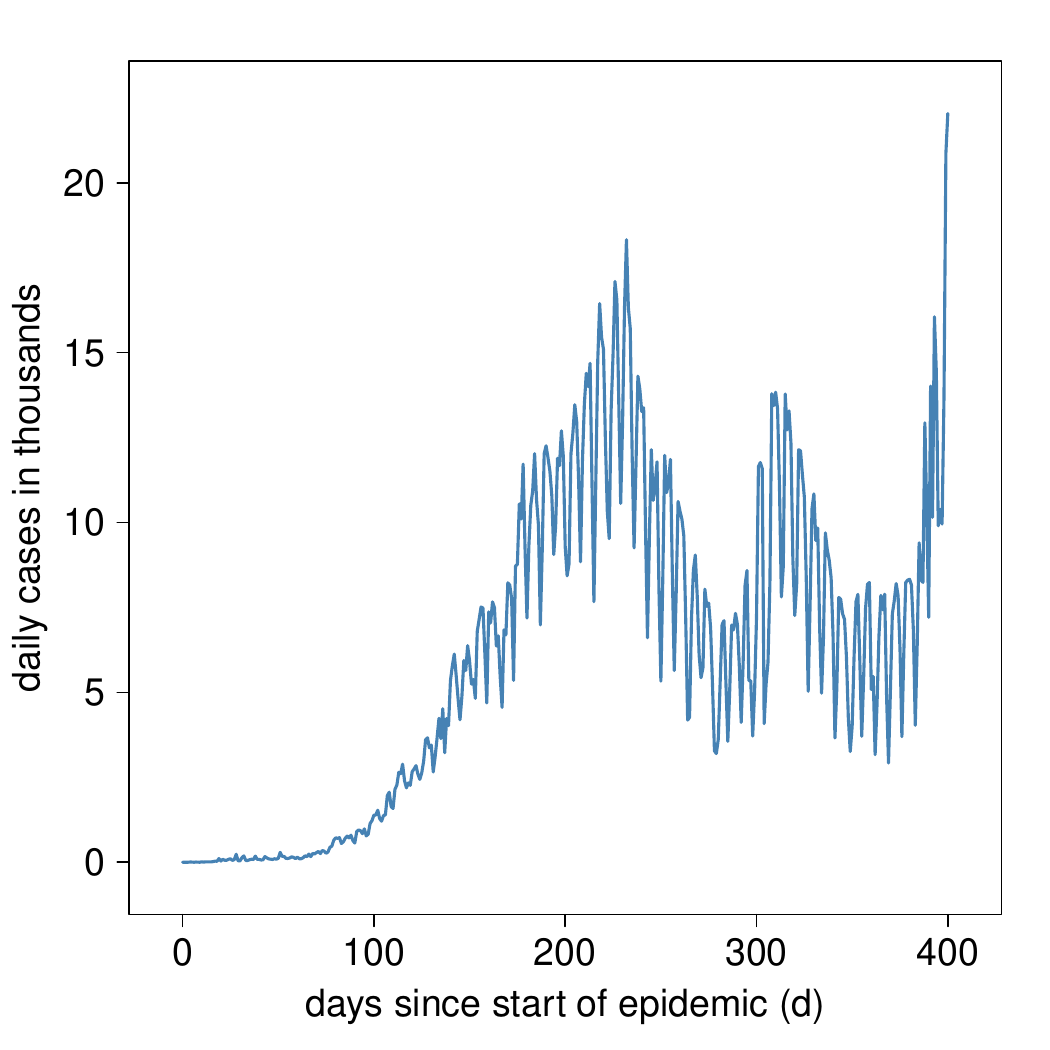}
	\end{subfigure}\\  
\caption{{Daily cases in thousands in India (left) and Argentina (right).}}
	\label{fig:casos}  
\end{figure}

\begin{figure}[H]
	\centering
	\includegraphics[width=0.5\linewidth]{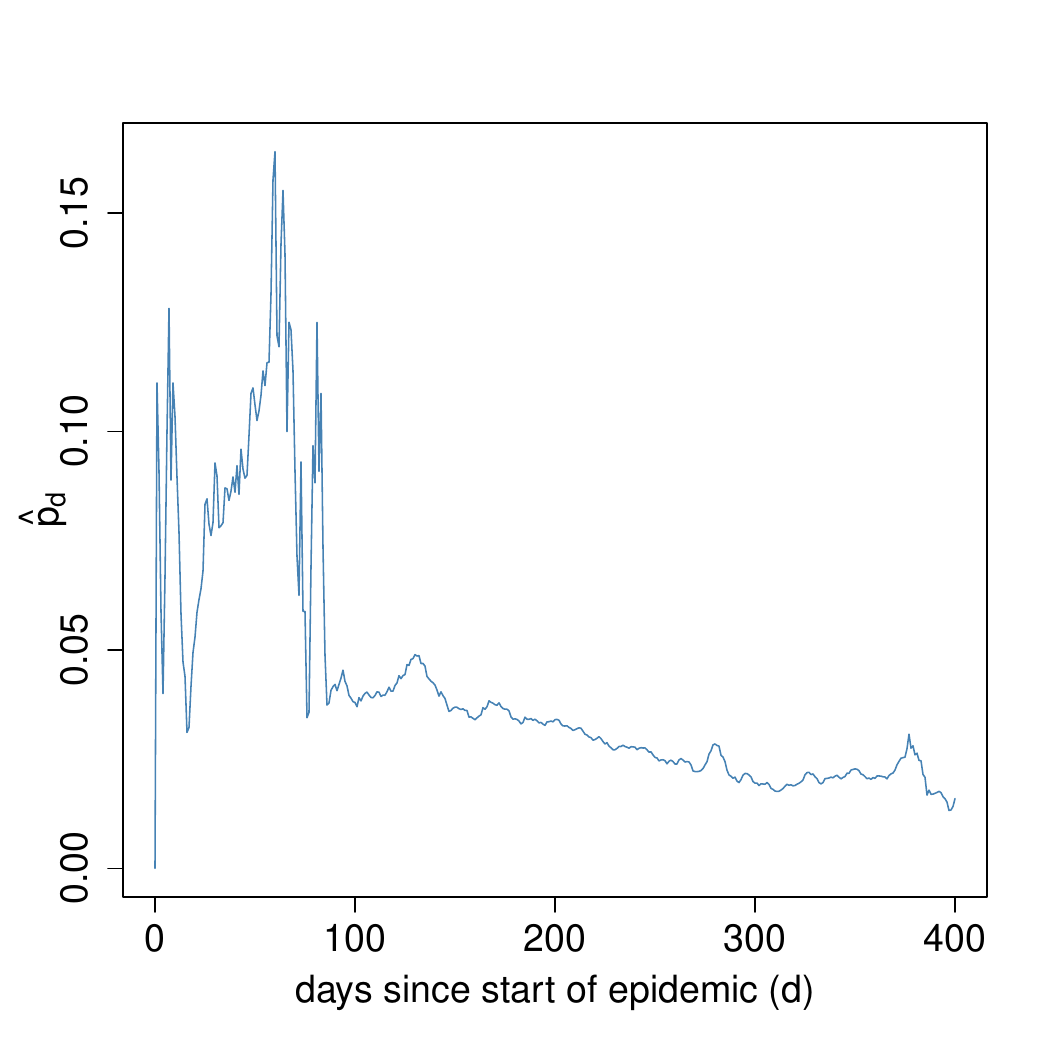}
	\caption{{Proportion of confirmed cases which finally died among those cases confirmed  between days $d-3$ and $d+3$} in the first $400$ days of the epidemic in Argentina}
	\label{fig:p_d}
\end{figure}

To evaluate the performance of the proposed estimators, a Monte Carlo study is performed, considering several hypothetical scenarios, varying the parameters $c_d$,  $p_d$ and the distribution function $F_d$. For $c_d$, values based on real data are chosen, specifically on the case of India and Argentina. In the first setting, the observed $c_d$ for $d=0$ to $d=400$ in Argentina are taken, while for the second, the observed $c_d$ in India for $d=36$ to $d=436$ are taken, since there were no observed cases from $d=1$ to $d=35$; see Figure \ref{fig:casos}. These quantities were downloaded from \verb|https://ourworldindata.org/coronavirus|; see \cite{owidcoronavirus}. For the values of $p_d$ two possibilities were considered. The first, that we will call Argentine $p_d$ is defined in the following way: 
for each day $d$, the value of $p_d$  is the proportion of confirmed cases which finally died among those cases confirmed  between days $d-3$ and $d+3$, \textsl{i.e}, during the week centered at $d$ in the Argentinian case; see Figure \ref{fig:p_d}. The second, that we will call the abruptly changing $p_d$, is defined as $p_d=0.05$ for $d=0$ to $120$ and $p_d=0.02$ for $d=121$ to $400$.
 Finally, two possible models for $F_d$ were considered. In both cases $F_d = F$  for all $d$. In the first model, called Argentine $F$, $F$ is a Zero Inflated Negative Binomial (ZINB) distribution, \textsl{i.e.}, a convex {combination} of  $F_0$, the distribution of a constantly zero variable, and $F_1$ a negative binomial distribution. We set $F=\pi F_0 + (1-\pi)F_1$, where $\pi=0.1$, and $F_1\sim  NB(\mu=12.6,r=1.2)$. This distribution has been seen to provide a very good fit in the real case of Argentina. Taking into account that a future epidemic might have different expected time from diagnosis to death,  a second $F_d$ equal to the Argentine $F$ except for  $\mu$, that changes to $\mu =6$ has been considered. This distribution function will be called Argentine $F$ with $\mu =6$.
In all cases $Nrep$ replicates of an epidemic developing form day $d=0$ to day $d=400$ have been simulated. For the Argentine $c_d$, $Nrep=1000$ is taken while for the Indian $c_d$, $Nrep =500$ is taken, to keep  computational times moderate,  since in this case $c_d$ are much larger.

 In all cases all the estimators  presented in Section \ref{sec:estimators} are calculated, namely $CFR_N(t)$, $CFR(t)$ and $CFR_G(t)$, as defined in equations \eqref{eq:naive}, \eqref{eq:cfrnuestra} and \eqref{eq:garske}, respectively.  $CFR(t)$ and $CFR_G(t)$ are calculated in two different ways. First, the estimation is made using the known distribution of survival times $F$ used to generate the data, and in such case the estimationis made  for  times $t$ with $10\le t\le 400$.  Second, the distribution $F$ is estimated by the ``empirical estimation'' $\hat F_{\textsc{emp}}$ in the way described in {Section \ref{sec:F}}. 
  Note that for the empirical estimation of $F$ on day $t$, in order to reduce bias,  
   it is needed to analyze the survival times of cases confirmed several days before day $t$ (until $t-t_{back}$); see {Section \ref{sec:F}}. 
   It is also necessary to have a relatively large number of deaths among confirmed cases in order to obtain a stable estimate of $F$. This is why, in this second instance, there is a need to begin the estimations a bit later, specifically on day $t_0=t_1+t_{back}$, where $t_1$ represents the days required to accumulate enough deaths by day $t_1+t_{back}$ from cases confirmed during the initial $t_1$ days. This is necessary for the estimation of $F$ to be stable, and $t_{back}$ is chosen such that $F(t_{back}) \ge 0.98.$
 In the cases of Argentine $F$ with $\mu=12.6$, we set $t_{back}=45$ and in the cases of $\mu=6$, we set $t_{back}=23$.  The values of $t_1$ were chosen as follows: in the cases of Argentine $c_d$, we took $t_1=45$ and in the cases of Indian $c_d$, since the number of cases is larger, we took $t_1=30$. In order to find a lower bound for $t_1$, note that, for $CFR(t_0)$ to be well defined in \eqref{eq:cfrnuestra}, it is necessary that $F_d(t_0-d)>0$ for all $d$  with $0\leq d\leq t_0$, and this holds if $F_d(0)>0$ for all $d$. When replacing $F_d$ by  $\hat F_{\textsc{emp}}$, the need is that $\hat F_{\textsc{emp}}(0)>0$. So $CFR(t_0)$ can be estimated only if, among cases confirmed between day $0$ and day $t_0-t_{back}=t_1$, at least one died the same day it was confirmed. 
      
We calculated $95\%$  confidence intervals for $cfr(t)$ presented in Section \ref{sec:estimators}: $CFR(t)\pm z_{0.975}\sqrt{\mathbb{V}(CFR(t))}$. Since $\mathbb{V}(CFR(t))$ depends both on the daily case fatality rates $p_d$ and on the distribution functions $F_d$  ($0\le d\le t$), when estimating $cfr(t)$ using the known $F$, also the known $F$ is used to estimate the variance whereas, when using, this estimation is also used to estimate $\mathbb{V}(CFR(t))$.
In both cases the estimation of the daily $p_d$ presented in \eqref{semanal} is performed. For the derivation of the confidence intervals, see the Appendix.

\begin{figure}[H]
	\centering
	\begin{subfigure}{0.5\textwidth}
		\centering
		\includegraphics[width=1\linewidth]{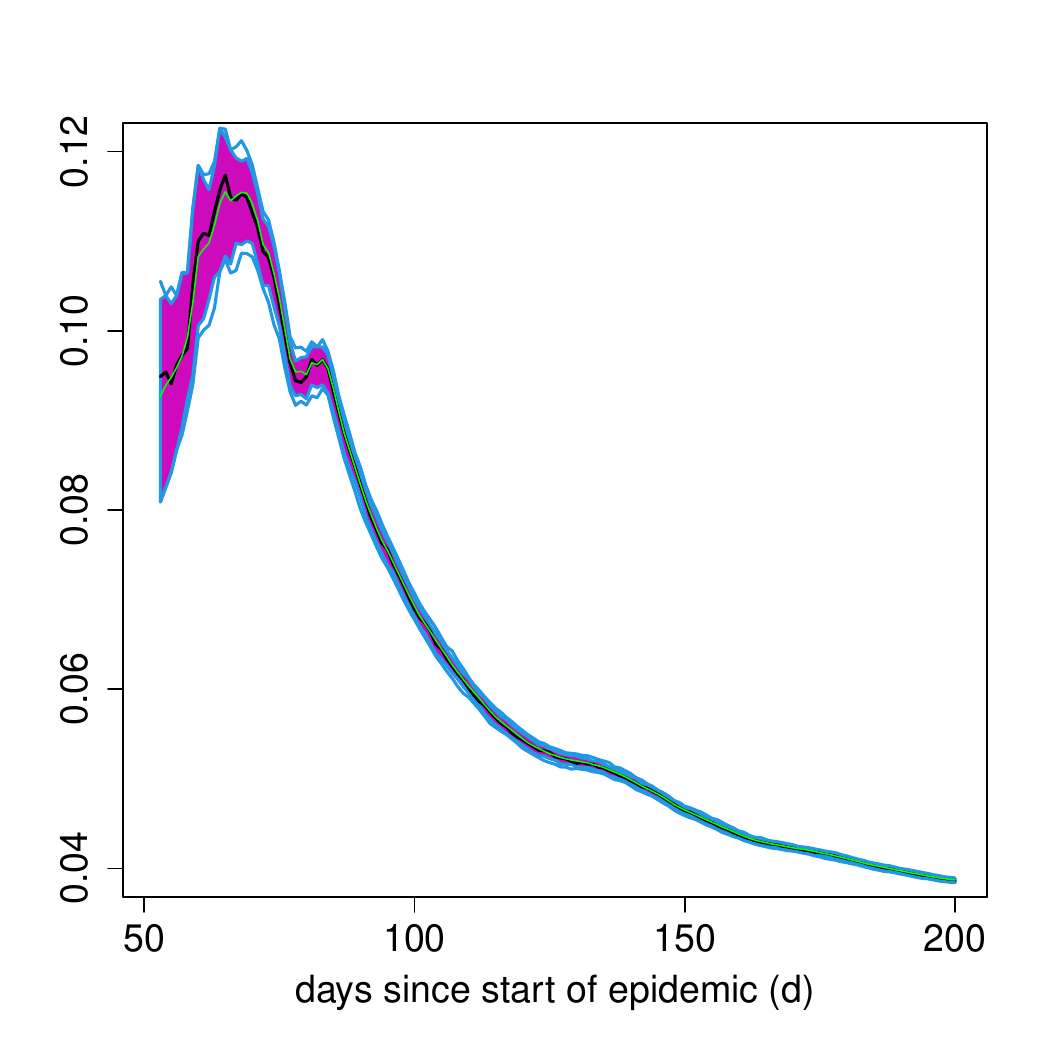}
	\end{subfigure}%
	\begin{subfigure}{0.5\textwidth}
		\centering
		\includegraphics[width=1\linewidth]{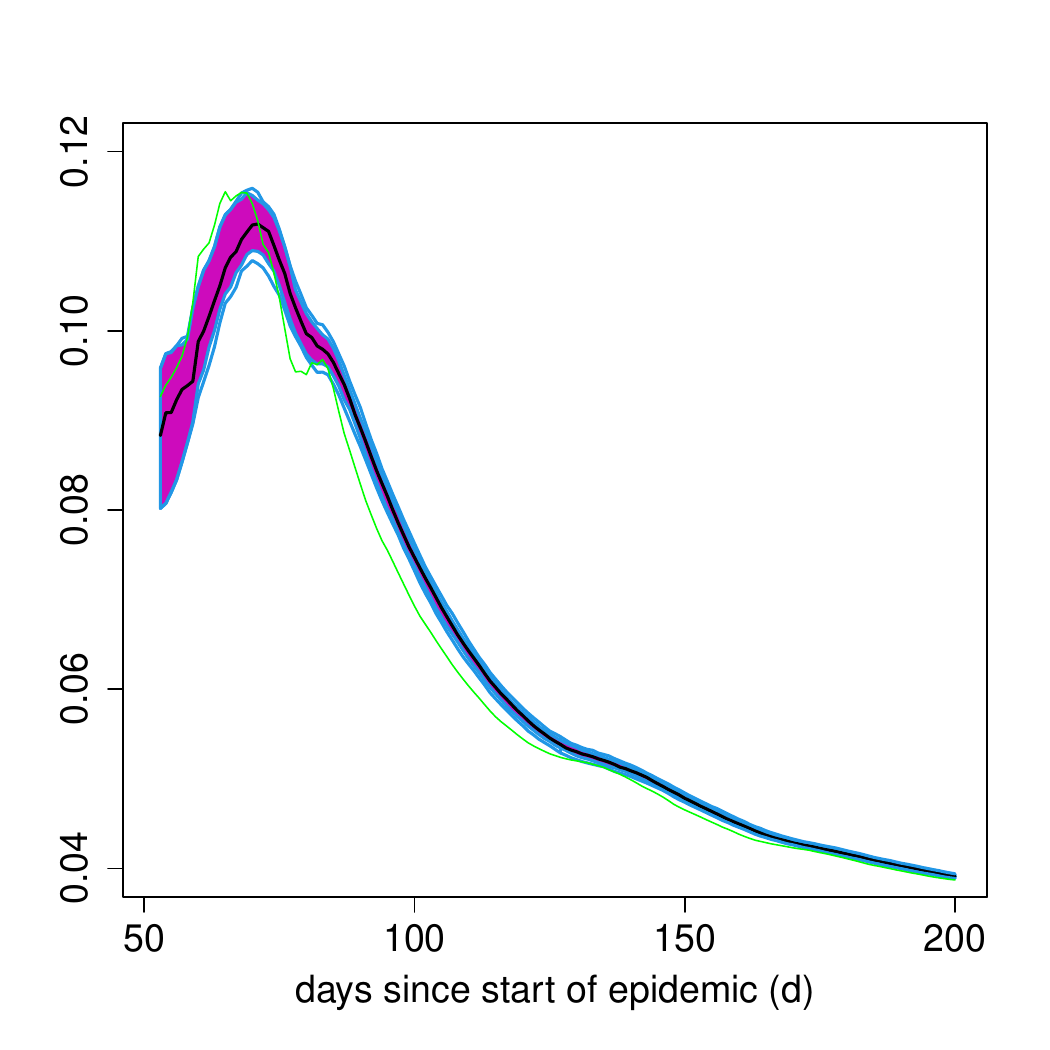}
	\end{subfigure}\\  
	\begin{subfigure}{0.5\textwidth}
		\centering
		\includegraphics[width=1\linewidth]{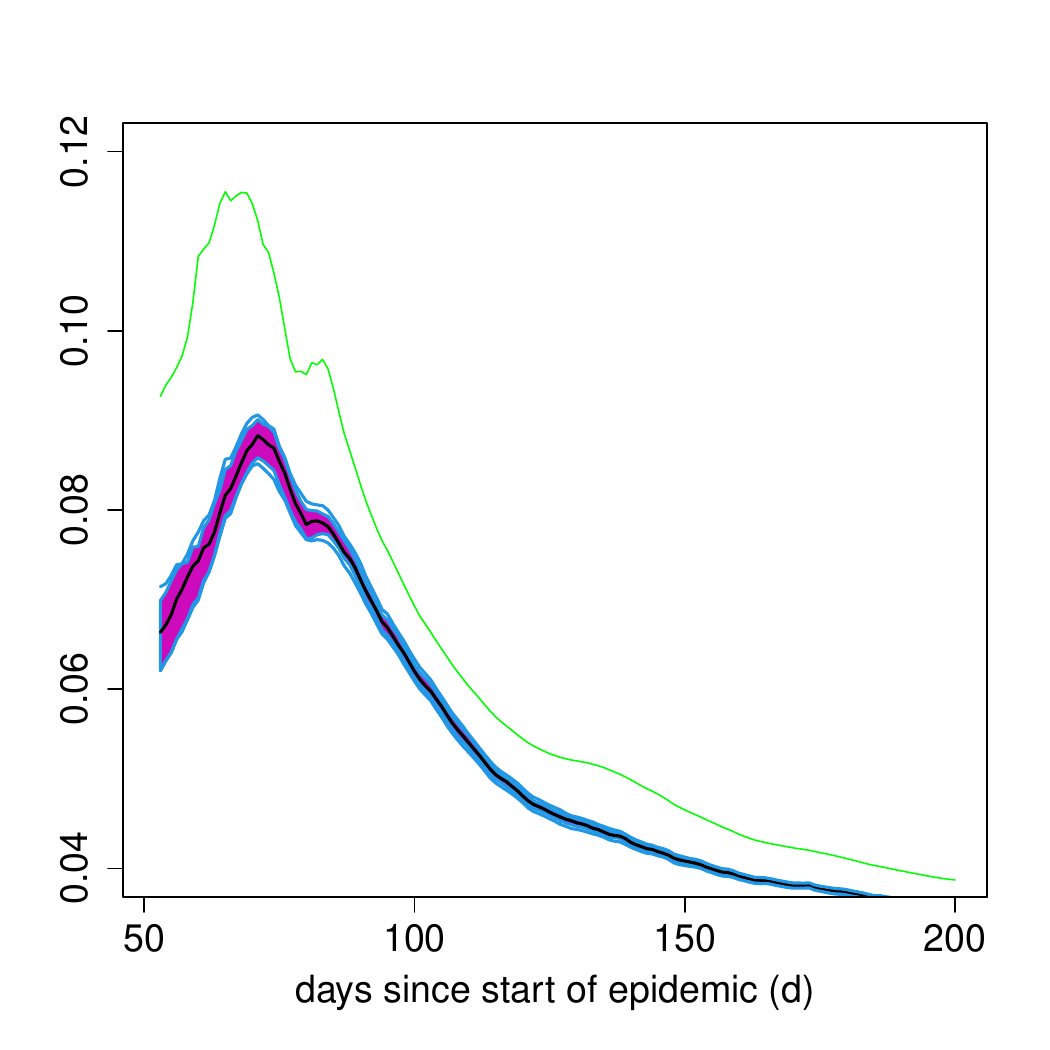}
	\end{subfigure}%
	\begin{subfigure}{0.5\textwidth}
		\centering
		\includegraphics[width=1\linewidth]{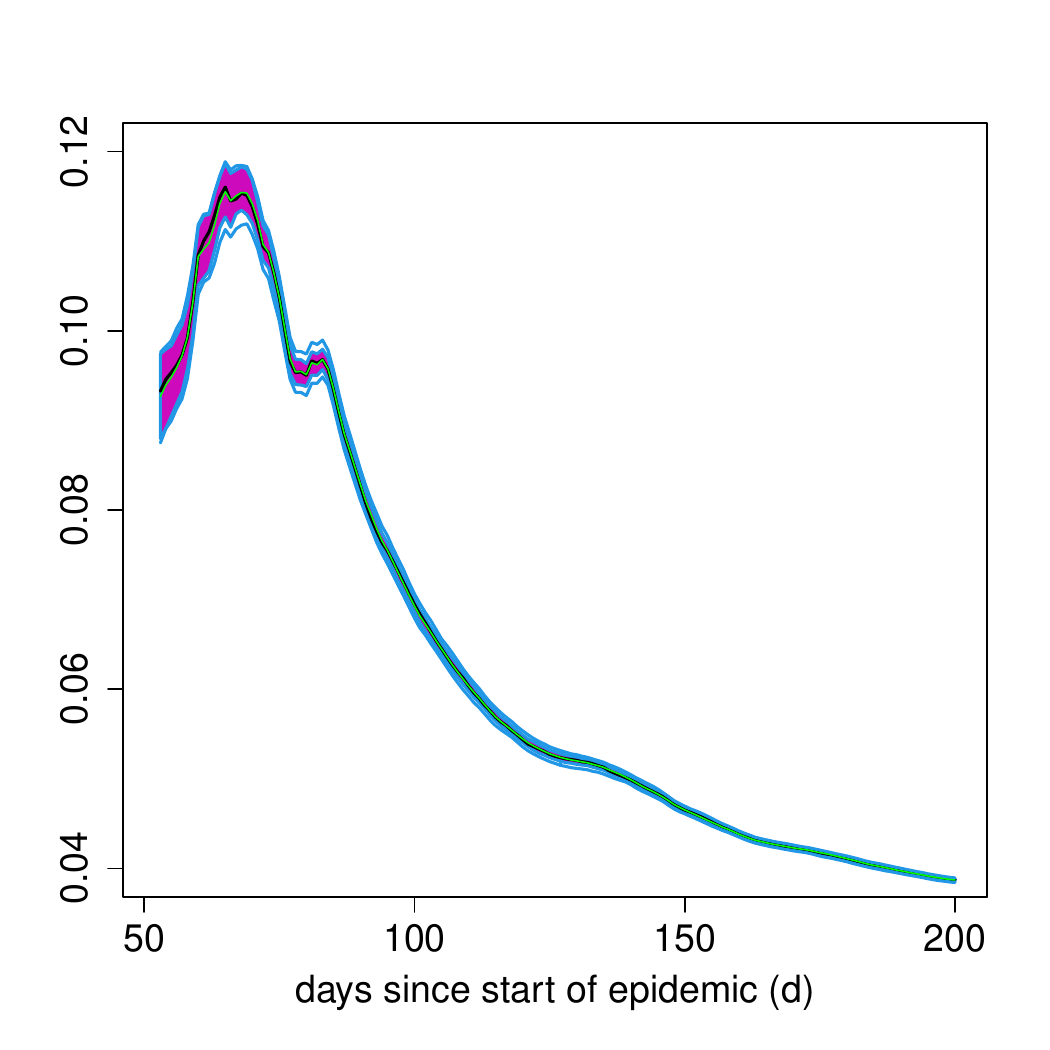}
	\end{subfigure}%
\caption{Functional boxplots of $CFR(t)$ (top left), $CFR_G(t)$ (top right), $CFR_N(t)$ (bottom left) and $CFR_F(t)$  (bottom right). The parameters used are: Indian $c_d$, $\mu=6$ and Argentine $p_d$. The estimation is made using $\hat F_{\textsc{emp}} $ with $t_1=30$ and $t_{back}=23$.}
	\label{fig:fboxplots2}  
\end{figure}

\begin{figure}[H]
	\centering
	\begin{subfigure}{0.5\textwidth}
		\centering
		\includegraphics[width=0.9\linewidth]{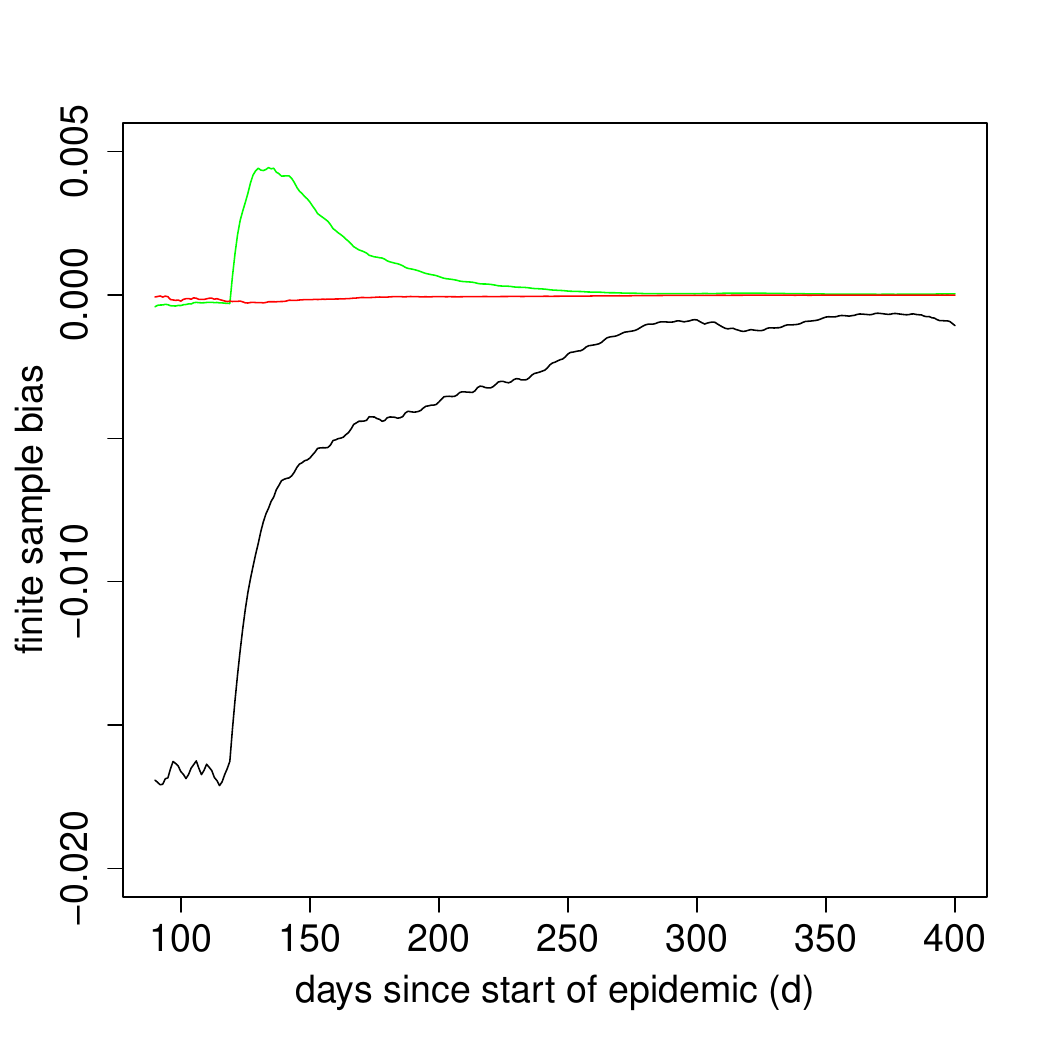}
	\end{subfigure}%
	\begin{subfigure}{0.5\textwidth}
		\centering
		\includegraphics[width=0.9\linewidth]{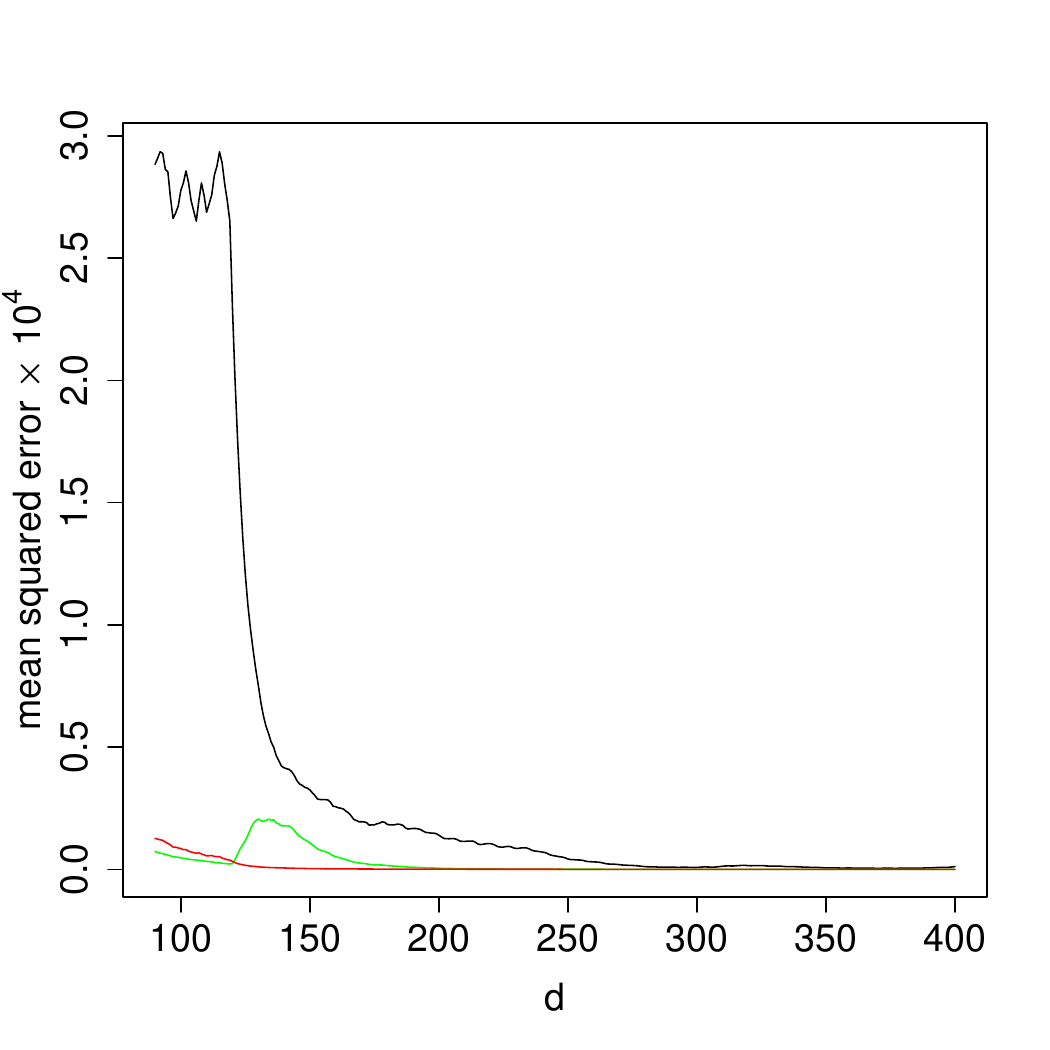}
	\end{subfigure}\\  
	\caption{Finite sample bias (left) and mean squared error multiplied by $10^4$ (right). Black curve corresponds to  $CFR_N(t)$, red curve to $CFR(t)$ and green curve to $CFR_G(t)$. 
		The parameters used are: Argentine $c_d$, $\mu=12.6$, abrupt $p_d$. The estimation is made using  $\hat F_{emp}$ with $t_1=45$ and $t_{back}=45$.   }
	\label{fig:finitesamplebiasmse1}
\end{figure}

\begin{figure}[H]
	\centering
	\begin{subfigure}{0.5\textwidth}
		\centering
		\includegraphics[width=0.9\linewidth]{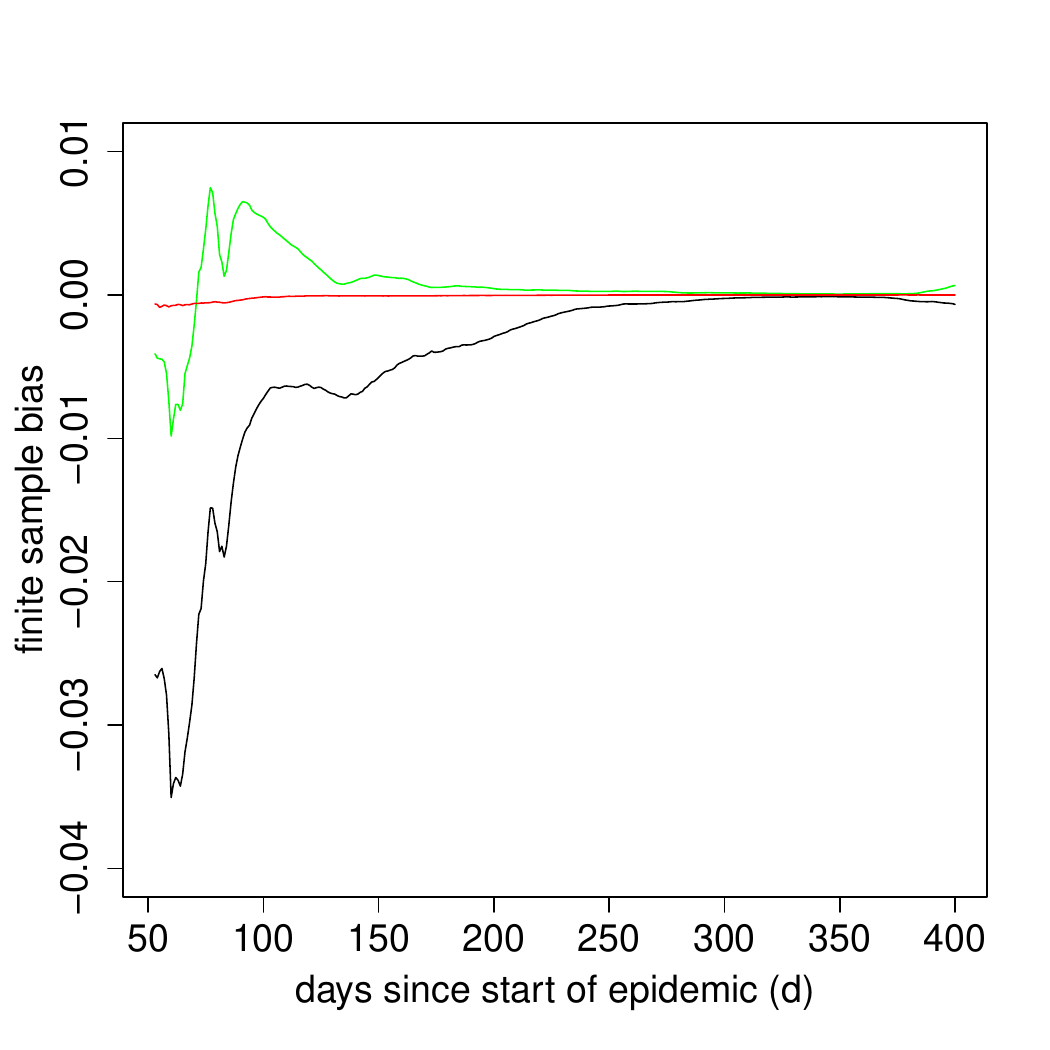}
	\end{subfigure}%
	\begin{subfigure}{0.5\textwidth}
		\centering
		\includegraphics[width=0.9\linewidth]{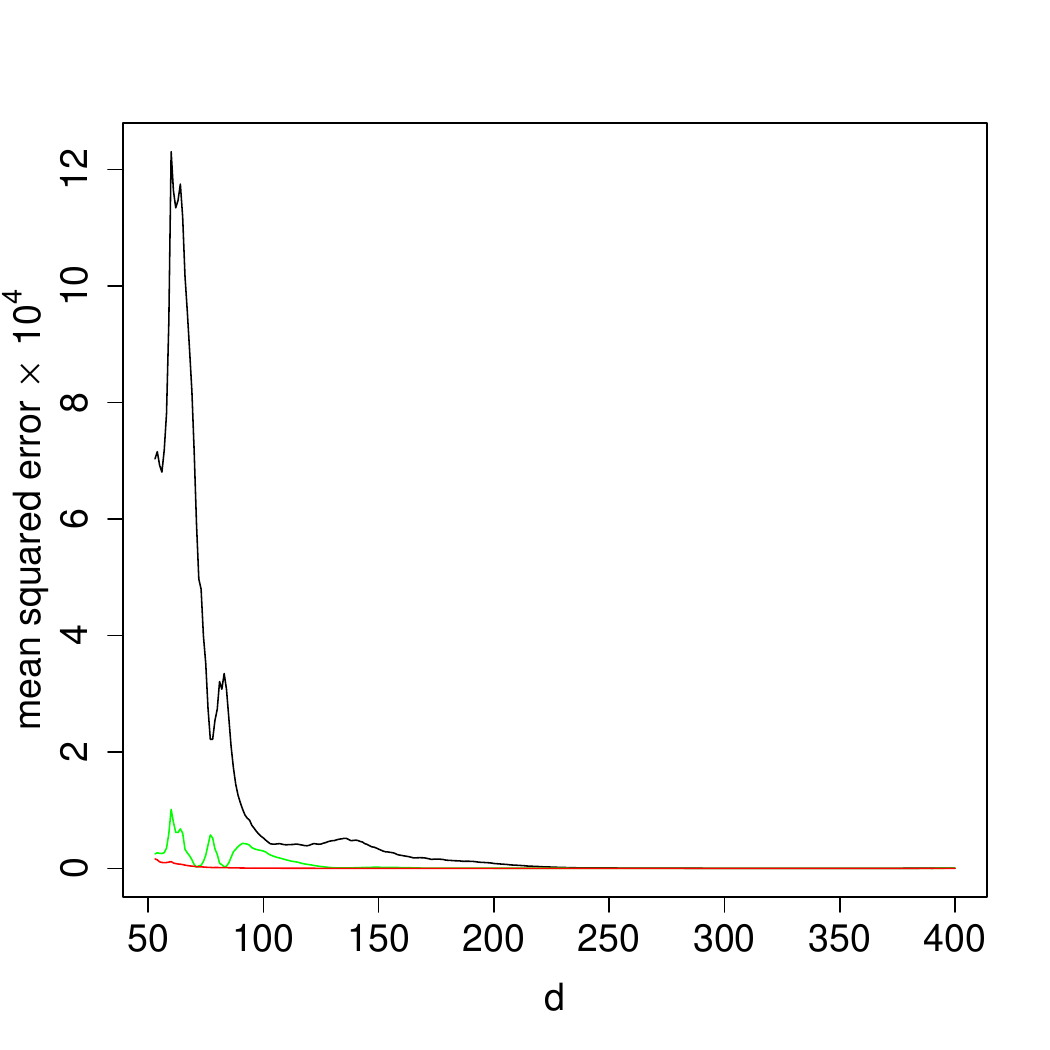}
	\end{subfigure}\\  
	\caption{Finite sample bias (left) and mean squared error multiplied by $10^4$ (right). Black curve corresponds to  $CFR_N(t)$, red curve to $CFR(t)$ and green curve to $CFR_G(t)$. 
		The parameters used are: Indian $c_d$, $\mu=6$ and Argentine $p_d$. The estimation is made using $\hat F_{emp}$ with $t_1=30$ and $t_{back}=23$.}
	
	\label{fig:finitesamplebiasmse2}
\end{figure}

\begin{figure}[H]
	\centering
	\begin{subfigure}{0.5\textwidth}
		\centering
		\includegraphics[width=1\linewidth]{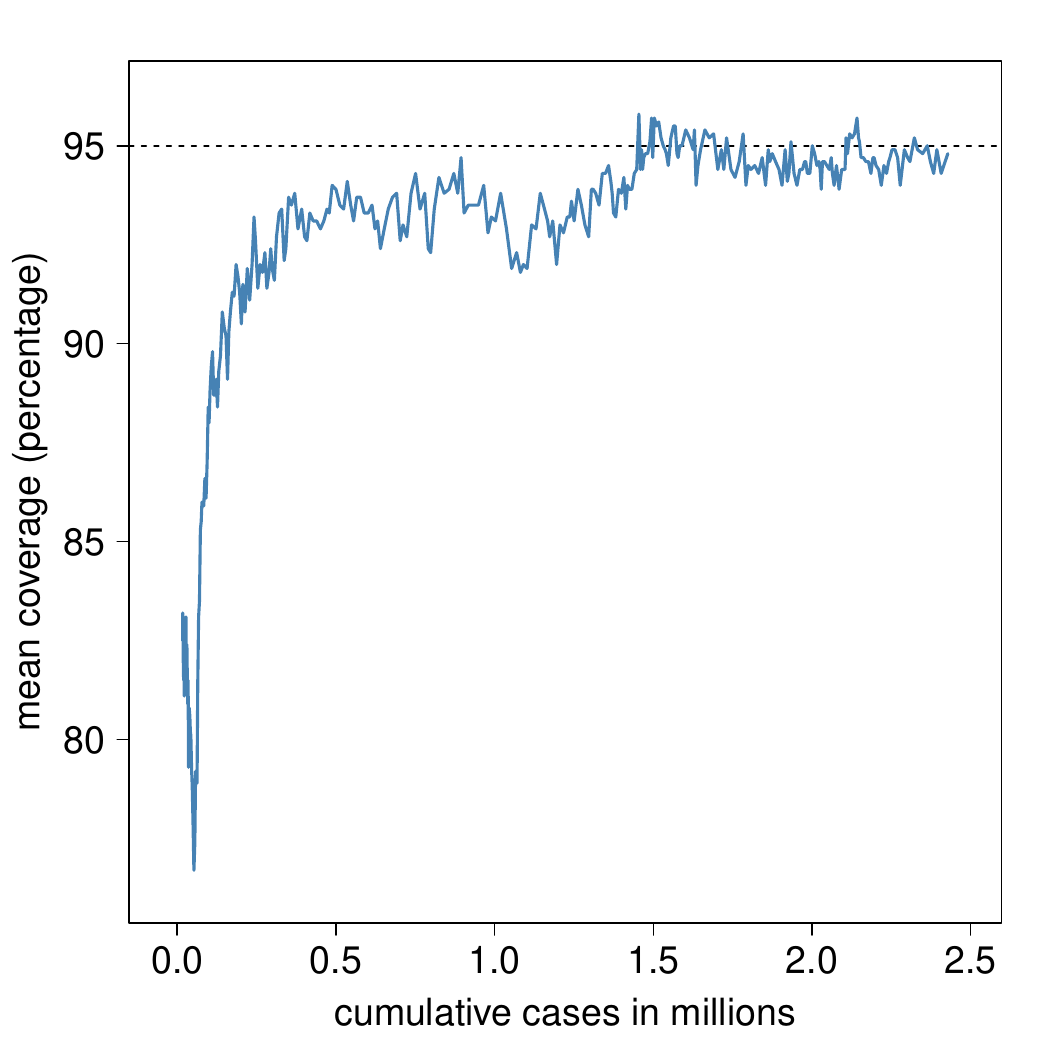}
	\end{subfigure}%
	\begin{subfigure}{0.5\textwidth}
		\centering
		\includegraphics[width=1\linewidth]{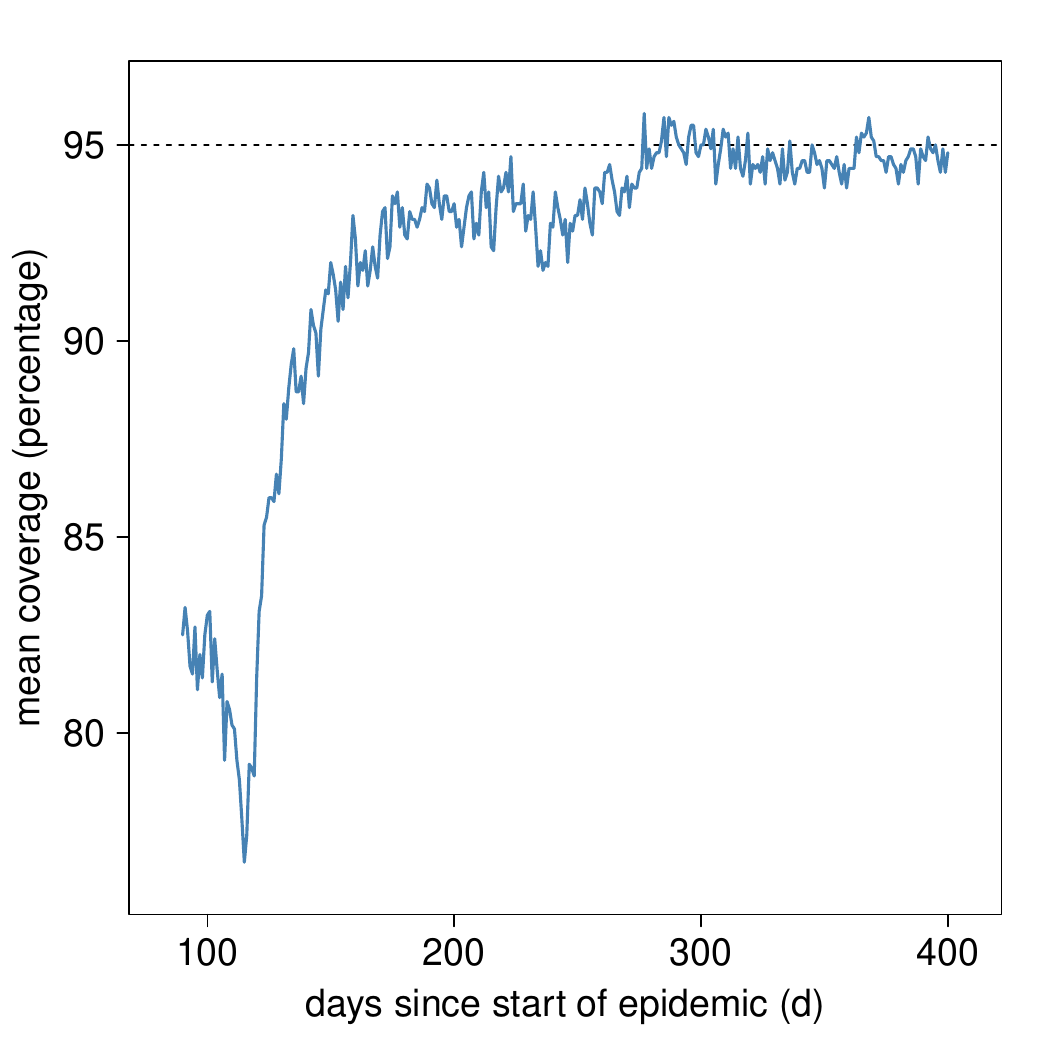}
	\end{subfigure}%
	\caption{Mean coverage of $95\%$ confidence intervals setting Argentine $c_d$, $\mu=12.6$, abrupt $p_d$, using $\hat F_{\textsc{emp}} $ with $t_1=45$ and $t_{back}=45$.}
	\label{fig:cubrimiento1}  
\end{figure}

\begin{figure}[H]
	\centering
	\begin{subfigure}{0.5\textwidth}
		\centering
		\includegraphics[width=1\linewidth]{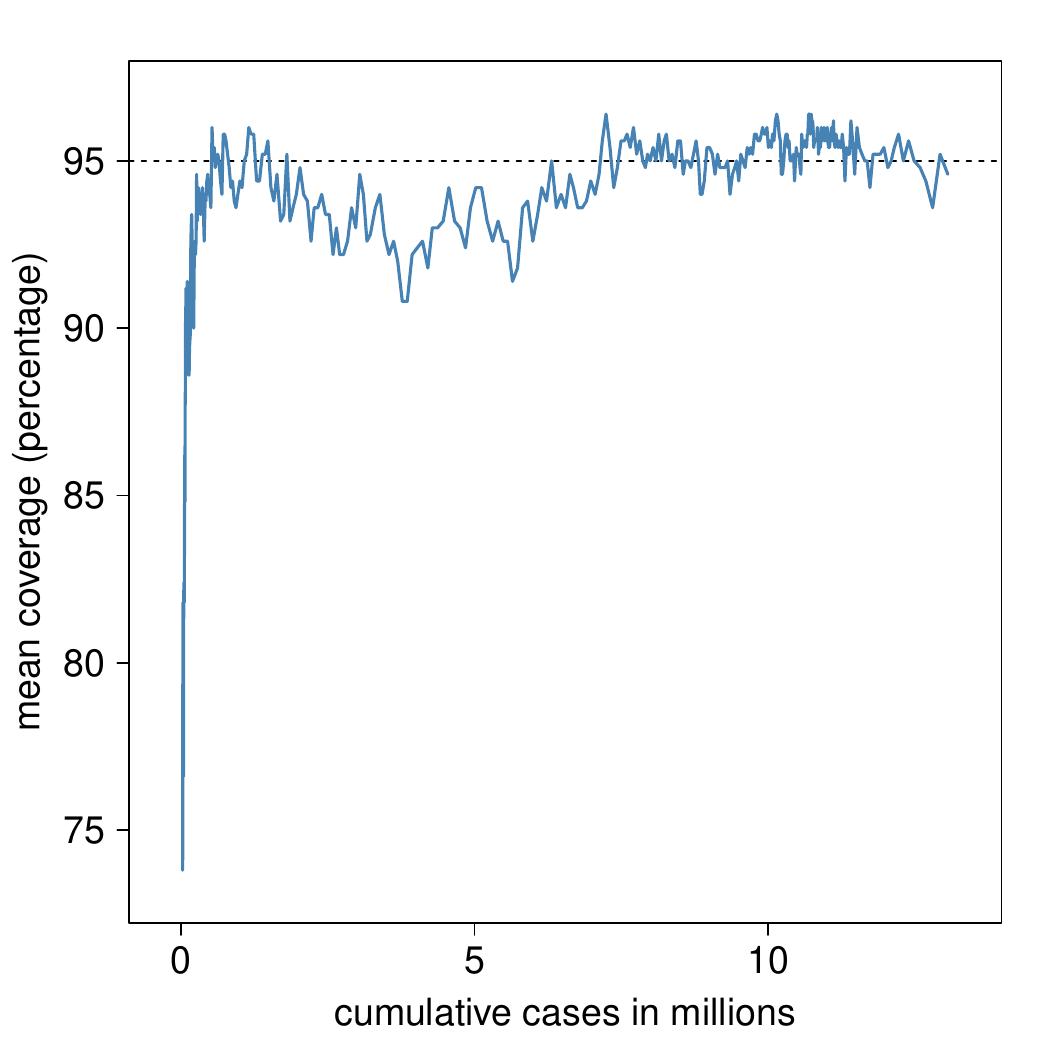}
	\end{subfigure}%
	\begin{subfigure}{0.5\textwidth}
		\centering
		\includegraphics[width=1\linewidth]{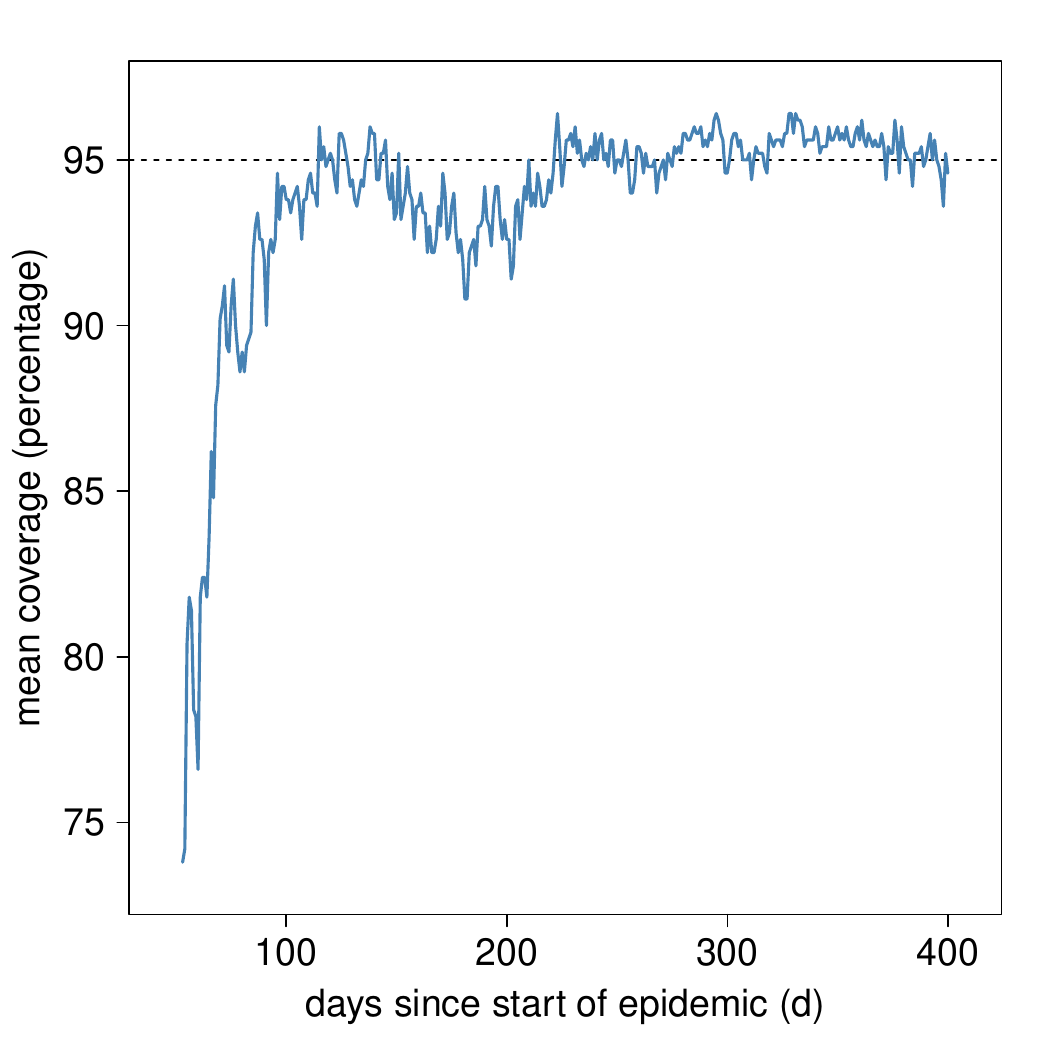}
	\end{subfigure}%
	\caption{Mean coverage of $95\%$ confidence intervals setting Indian $c_d$, $\mu=6$ and Argentine $p_d$, using $\hat F_{\textsc{emp}} $ with $t_1=30$ and $t_{back}=23$.}
	\label{fig:cubrimiento2}  
\end{figure}

Due to space limitations,  in Figures \ref{fig:fboxplots1} to \ref{fig:cubrimiento2}, the graphics to analyze the simulation results for only two of the considered scenarios are displayed, namely: Argentine $c_d$, $\mu=12.6$, abrupt $p_d$ and $F$ estimated by $\hat F_{\textsc{emp}} $ with $t_1=45$ and $t_{back}=45$; and Indian $c_d$, $\mu=6$, Argentine $p_d$ and $F$ estimated by $\hat F_{\textsc{emp}} $ with $t_1=30$ and $t_{back}=23$.

In order to compare  $CFR_N(t)$, $CFR(t)$ and $CFR_G(t)$ with $cfr(t)$, a functional boxplot of the $Nrep$ estimated curves obtained by each method is presented. Also the functional boxplot of $CFR_F(t)$ is presented, which is the best possible estimation of $cfr(t)$, but only computable at the end of the epidemic, as a reference.    In each functional boxplot  a plot of the curve to be estimated, $cfr(t)$, is added; see Figures \ref{fig:fboxplots1} and \ref{fig:fboxplots2}.

	\begin{figure}[H]
	\centering
	\begin{subfigure}{0.5\textwidth}
		\centering
		\includegraphics[width=1\linewidth]{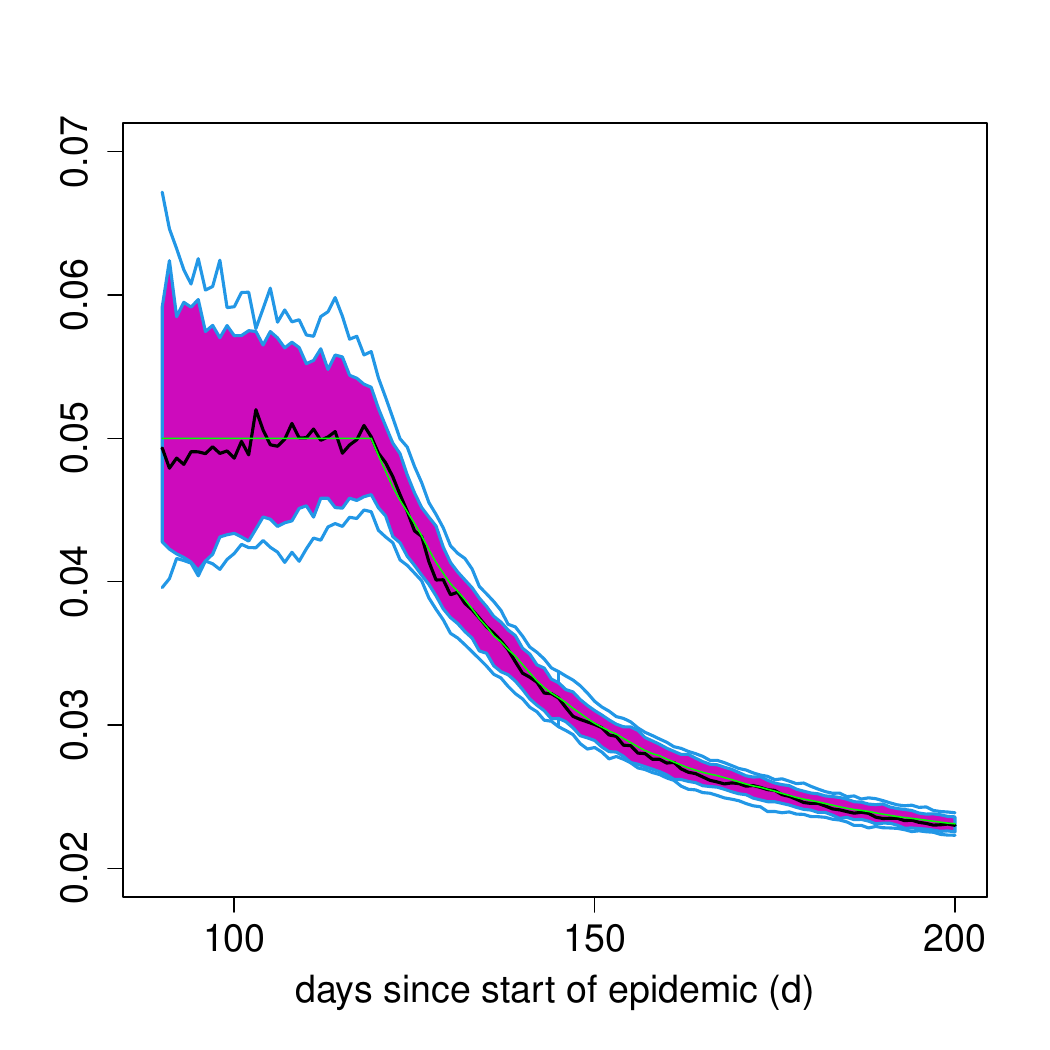}
	\end{subfigure}%
	\begin{subfigure}{0.5\textwidth}
		\centering
		\includegraphics[width=1\linewidth]{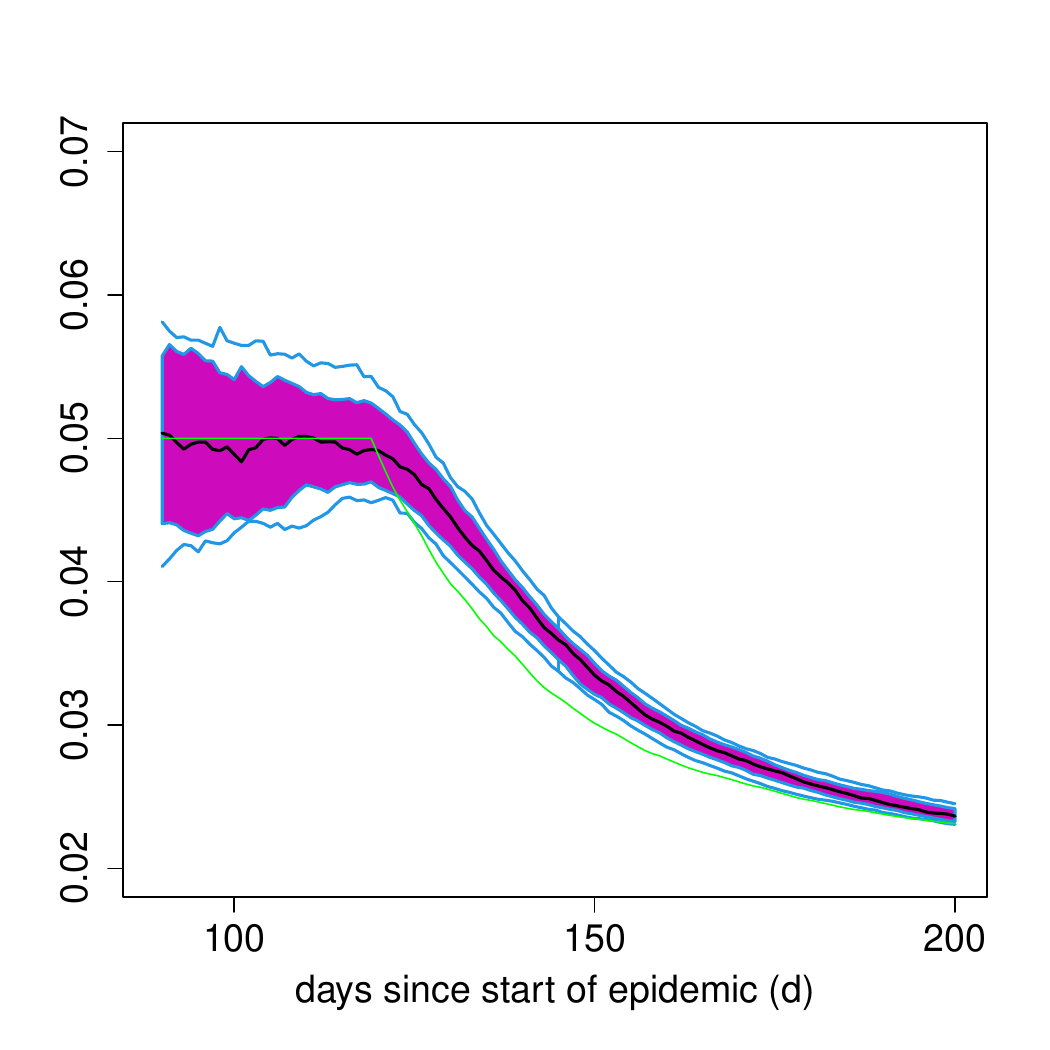}
	\end{subfigure}\\  
		\begin{subfigure}{0.5\textwidth}
		\centering
		\includegraphics[width=1\linewidth]{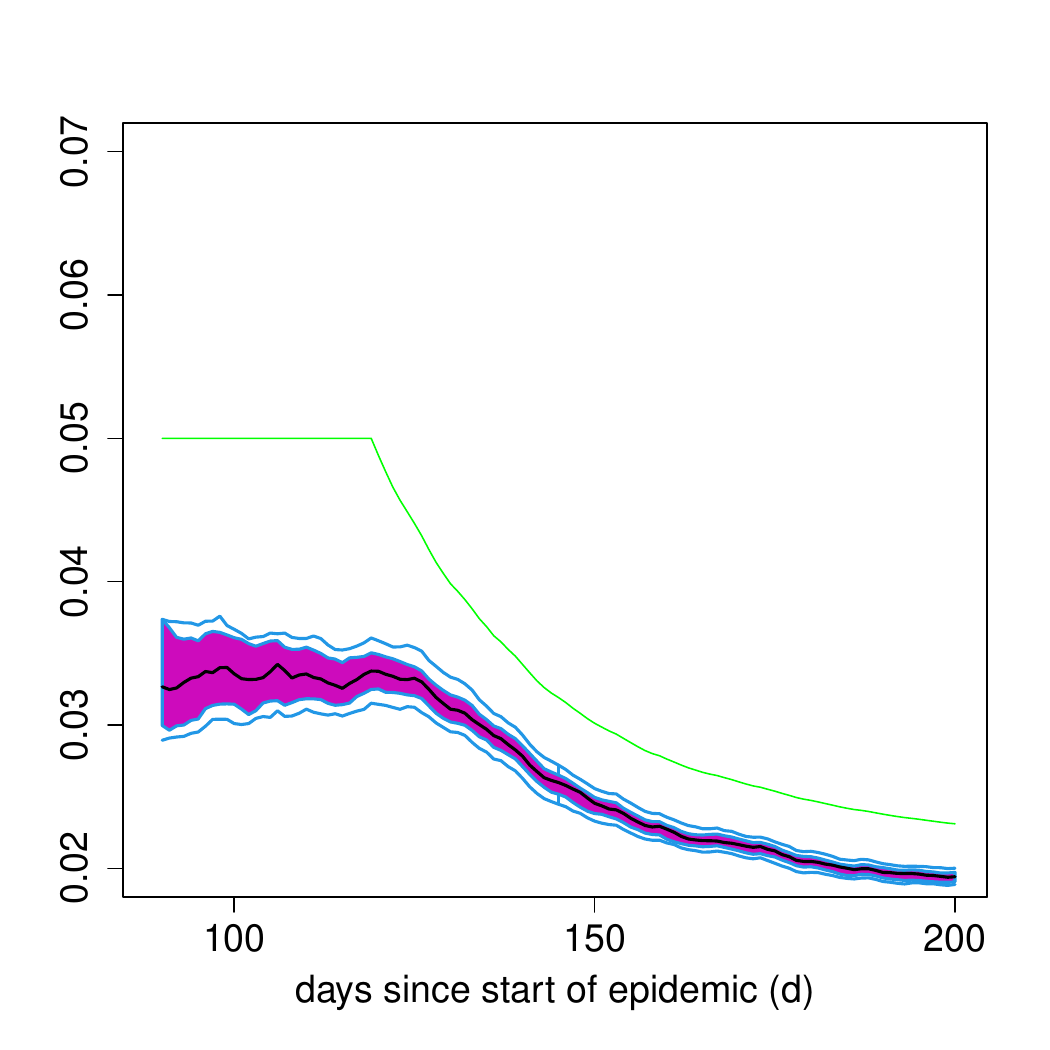}
	\end{subfigure}%
	\begin{subfigure}{0.5\textwidth}
	\centering
	\includegraphics[width=1\linewidth]{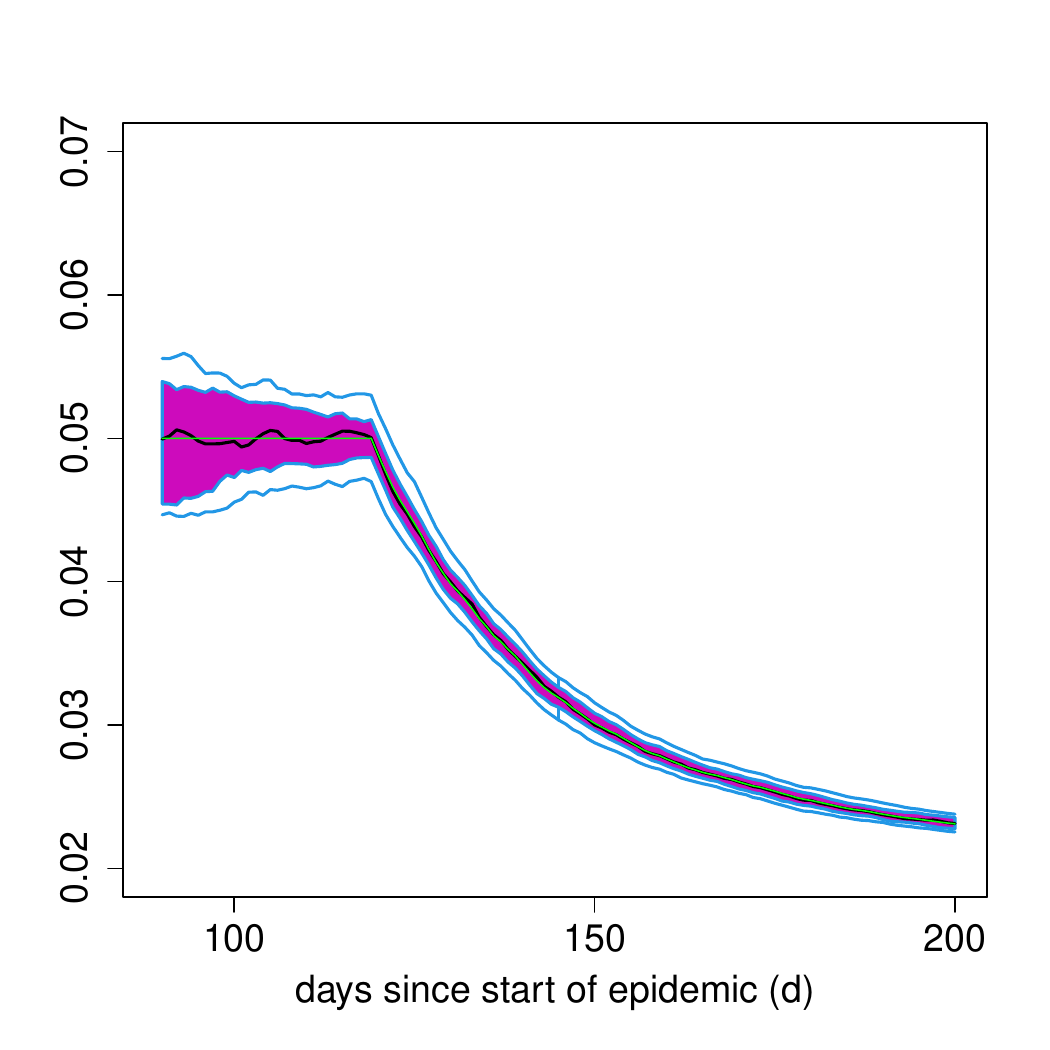}
\end{subfigure}%
\caption{Functional boxplots of $CFR(t)$ (top left), $CFR_G(t)$ (top right), $CFR_N(t)$ (bottom left) and $CFR_F(t)$  (bottom right). The parameters used are: Argentine $c_d$, $\mu=12.6$, abrupt $p_d$. The estimation is made using $\hat F_{\textsc{emp}} $ with $t_1=45$ and $t_{back}=45$.}
	\label{fig:fboxplots1}  
\end{figure}

In Figures \ref{fig:fboxplots1} and \ref{fig:fboxplots2}  only results until day 200 are shown in order to make  clearly visible the behaviour of the estimators at the beginning of the epidemic. 
In Figure \ref{fig:fboxplots1}, one can observe the unbiased behaviour of $CFR(t)$ during this period. On the other hand, the estimator proposed by \cite{garske2009assessing}, $CFR_G(t)$, overestimates  $cfr(t)$ for a long period of time after day $120$, the day when the cumulative  case fatality rates $cfr(t)$ start to decrease. On the other hand one observes the severe tendency to underestimate of the ``naive'' case fatality rate $CFR_N(t)$ during the outbreak of the epidemic. 

In Figure \ref{fig:fboxplots2} one can observe that $CFR_G(t)$ underestimates $cfr(t)$ in the period of time in which $cfr(t)$ is increasing and  overstimates $cfr(t)$ when it is decreasing.
 The unbiased behaviour of $CFR(t)$ is also observed and the severe tendency to underestimate of $CFR_N(t)$ during the outbreak of the epidemic. A similar behaviour between $CFR_F(t)$ and $CFR(t)$ is observed, except that, as may be expected, $CFR_F(t)$ has less variability.

Though it is not visible in Figures \ref{fig:fboxplots1} and \ref{fig:fboxplots2}, by day $400$, since the number of cases is very large, all the estimators give good results.

In Figures \ref{fig:finitesamplebiasmse1} and  \ref{fig:finitesamplebiasmse2} (below) both the finite sample bias and the finite sample Mean Squared Error of estimators for the same scenarios plotted in Figures \ref{fig:fboxplots1} and \ref{fig:fboxplots2}, respectively, are presented. 
In Figure \ref{fig:finitesamplebiasmse1}  a lower mean squared error of $CFR_G$ compared to $CFR$ is observed for the first period until day $120$, moment in which $cfr(t)$ starts to decrease and the bias of $CFR_G$ produces a higher mean squared error compared to $CFR$ as well.
In Figure \ref{fig:finitesamplebiasmse2} one can observe more clearly  how $CFR_G$ has periods of moderate negative bias and of moderate positive bias. These periods correspond to the increasing and decreasing periods of $cfr(t)$, respectively, showing that $CFR_G(t)$ has a delay at detecting changes in $cfr(t)$, whereas $CFR(t)$ reacts to these changes instantly. 

We have observed these same phenomena described for Figures \ref{fig:fboxplots1}, \ref{fig:fboxplots2}, \ref{fig:finitesamplebiasmse1} and \ref{fig:finitesamplebiasmse2} in all scenarios analyzed in the simulation, that is to say, minimum finite sample bias for $CFR(t)$ for all $t$, moderate finite sample bias for $CFR_G(t)$ in periods of changing $cfr(t)$ and even larger finite sample bias for $CFR_N(t)$ for a large period of time since the beginning of the epidemic.

In the Appendix  it is proved that the theoretical confidence intervals presented in Section \ref{sec:estimators} have an assympthotic level of $(1-\alpha)\times 100 \%$. However, the C.I. calculated in the simulations may differ slightly from the theoretical ones. In some cases empirical $F$ may be used instead of theoretical $F$, and the calculation of $\mathbb{V}(CFR(t))$
uses $\hat{p}_d$ defined in \eqref{semanal} instead of $p_d$. It is important to determine if these computable C.I.
also have an approximate confidence level of $(1-\alpha)\times 100 \%$  and to analyze how many confirmed cases are needed for the C.I. to reach the desired level of confidence. To address this, the C.I. have been calculated in all scenarios of the simulations. For each replication in every scenario, the confidence interval $(a(t), b(t))$ is constructed for $t_0\leq t\leq 400$. Then it is checked if each interval contains the corresponding $cfr(t)$. Finally, it is measured the proportion of times, out of the $Nrep$ replications, in which $a(t) \le cfr(t) \le b(t)$. We refer to this proportion as the empirical mean coverage of the C.I. at time $t$.

In Figures  \ref{fig:cubrimiento1} and  \ref{fig:cubrimiento2}  the empirical coverage of the C.I. is plotted as a function of the number of days since the beginning of the epidemic and as a function of the number of confirmed cases.
In Figure  \ref{fig:cubrimiento1} one  observes that the empirical mean coverage of the C.I. is arround $95\%$ approximately since day $270$ or for $1.5$ million of confirmed cases, and stabilizes around this empirical coverage from that moment onwards, while the empirical coverage exceeds $90\%$ with more than $250000$ confirmed cases. In Figure  \ref{fig:cubrimiento2}  one can observe that the empirical mean coverage of the C.I. reaches $95\%$ around day $120$ and stabilizes in that level of coverage approximately by day $220$.  In Tables \ref{tab:empcovempF} and \ref{tab:empcovknownF} the emprirical coverage for all the simulation scenarios considered is given. In Table \ref{tab:empcovempF}, the results when estimating $F$ using $\hat F_{\textsc{emp}} $ are presented. It is observed that in all cases, the level of empirical coverage stabilizes at $95\%$ at some point. In cases where $\mu=6$, this is achieved significantly sooner. In Table \ref{tab:empcovknownF}, the results with known $F$ are presented. One can observe that, in this case, the desired coverage of $95\%$ is reached approximately around day $25$.


\begin{table}[ht]\small
\centering
\begin{tabular}{llll|rrrrrrrr}
\hline
 &&&& \multicolumn{8}{|c}{Days since the beginning of the epidemic}\\
  \hline
$c_d$ & $p_d$ & $\mu$ &$t_{back}$ & 75 & 100 & 150 & 200 & 250 & 300 & 350 & 400 \\ 
  \hline
  India & Abrupt & 12.6 & 45 &  52.2 & 72.0 & 85.0 & 88.2 & 92.8 & 93.4 & 93.6 & 94.8 \\ 
   India & Abrupt & 6 & 23 & 91.8 & 91.8 & 93.6 & 94.0 & 94.4 & 95.2 & 93.8 & 94.4 \\  
  India & Arg & 12.6 & 45 & - & 72.4 & 88.6 & 72.4 & 90.2 & 95.2 & 95.8 & 94.4 \\ 
  India & Arg & 6 & 23 & 90.6 & 93.8 & 93.4 & 92.6 & 95.0 & 94.6 & 96.0 & 94.6 \\ 
  Arg & Abrupt & 12.6 & 45 & - & 83.0 & 92.0 & 93.5 & 93.2 & 95.0 & 93.9 & 94.8 \\
    Arg & Abrupt & 6 & 23 & 95.2 & 91.6 & 93.8 & 94.3 & 93.9 & 95.5 & 95.0 & 95.2 \\  
  Arg & Arg & 12.6 & 45 & - & 90.0 & 91.8 & 91.5 & 93.4 & 94.6 & 95.7 & 94.9 \\ 
  Arg & Arg & 6 & 23 & 96.0 & 95.8 & 94.2 & 94.8 & 95.2 & 94.5 & 95.3 & 95.2 \\ 
   \hline
\end{tabular}\caption{Empirical coverage of confidence intervals computed with $\hat F_{\textsc{emp}} $.}\label{tab:empcovempF}
\end{table}

\begin{table}[ht]\small
\centering
\begin{tabular}{lll|rrrrrrrrrr}
\hline
 &&& \multicolumn{8}{|c}{Days since the beginning of the epidemic}\\
  \hline
$c_d$ & $p_d$ & $\mu$ & 25 & 50 & 75 & 100 & 150 & 200 & 250 & 300 & 350 & 400 \\ 
  \hline
India & Abrupt & 12.6 & 95.4 & 95.7 & 95.6 & 96.4 & 96.0 & 95.8 & 95.1 & 93.9 & 94.1 & 94.7 \\ 
  India & Abrupt & 6 & 96.6 & 95.6 & 95.3 & 95.6 & 95.0 & 95.5 & 95.3 & 94.9 & 94.7 & 94.1 \\ 
   India & Arg & 12.6 & 94.1 & 95.1 & 96.6 & 95.9 & 95.5 & 95.7 & 94.5 & 95.1 & 95.6 & 94.7 \\ 
  India & Arg & 6 & 94.0 & 95.8 & 96.3 & 95.6 & 95.7 & 95.2 & 95.6 & 94.8 & 96.0 & 94.8 \\ 
  Arg & Abrupt & 12.6 & 95.4 & 94.9 & 96.2 & 95.5 & 95.3 & 95.0 & 94.7 & 95.2 & 94.1 & 94.9 \\ 
  Arg & Abrupt & 6 & 95.5 & 95.8 & 96.3 & 95.5 & 95.2 & 95.0 & 94.0 & 95.4 & 94.8 & 95.3 \\ 
    Arg & Arg & 12.6 & 92.9 & 95.4 & 97.0 & 96.0 & 96.2 & 95.1 & 95.9 & 95.0 & 95.5 & 94.8 \\ 
  Arg & Arg & 6 & 94.8 & 94.8 & 96.5 & 96.4 & 95.5 & 94.6 & 95.5 & 94.6 & 95.4 & 95.3 \\ 
   \hline
\end{tabular}\caption{Empirical coverage of confidence intervals computed with known $F$.}\label{tab:empcovknownF}
\end{table}

\section{Real data: the COVID-19 epidemic in \\Argentina}\label{sec:datosReales}
	{In this section, the behavior of the estimators presented in Section \ref{sec:estimators} is analyzed in a real data example. Of course, in  this case,  the values of the estimators can not be compared with the values of interest $cfr(t)$, as done in the simulation study, because these values are not observable. The natural way to deal with this problem is to compare the estimators to the final case fatality rate by day $t$, $CFR_F(t)$.}

For each day $t$ from June 1st to December 31st 2020, different estimators of $cfr(t)$ are computed using the data base from the Ministry of Health of Argentina  as of April 4th, 2021. 
See Appendix \ref{app2} for a description of the data.
Three estimators of the case fatality rate by day $t$,  $cfr(t)$, that can be computed on day $t$,  namely $CFR_N(t)$, $CFR(t)$ and $CFR_G(t)$ are compared to  the final case fatality rate until day $t$ observed on April 4th 2021, {which we  call $CFR_F(t)$, assuming that all the confirmed cases during 2020 have either recovered or died by that date}. The three estimators are defined in equations \eqref{eq:naive}, \eqref{eq:cfrnuestra} and \eqref{eq:garske}, respectively, and $CFR_F(t)$ is computed by
	\begin{equation}\label{finalTrucho}
CFR_F(t) = \frac{\displaystyle\sum_{d=0}^t \displaystyle\sum_{i=1}^{c_d} D_{d,i}(t_F)} {\displaystyle\sum_{d=0}^t c_d}
	\end{equation}
where $t_F$ is the number of days from March 3rd 2020 to April 4th	2021. {Observe that $CFR_F(t)$ defined in \eqref{finalTrucho} equals the one defined in \eqref{final} if all people diagnosed until day $t$ have already died or recovered by day $t_F$, which might not be true but both values should be very close one from the other. In fact,}
 the Argentine data set shows that the $0.98$ quantile of the times between confirmation and death is $45$ days. We also observe that the ditributions $F_d$ defined in Section \ref{sec:defi} do not change significantly when $d$ varies. 
 For this reason, we assume $F_d \equiv F$ and estimate it by $\hat  F_{\textsc{emp}} $,  using $t_{back}=45$ and $t_1=45$; see Section \ref{sec:F} for the definitions of  $\hat  F_{\textsc{emp}} $,  and $t_{back}$ and  Section \ref{sec:montecarlo} for the definition of $t_1$.
	
\begin{figure}[H]
		\centering
		\includegraphics[width=0.7\linewidth]{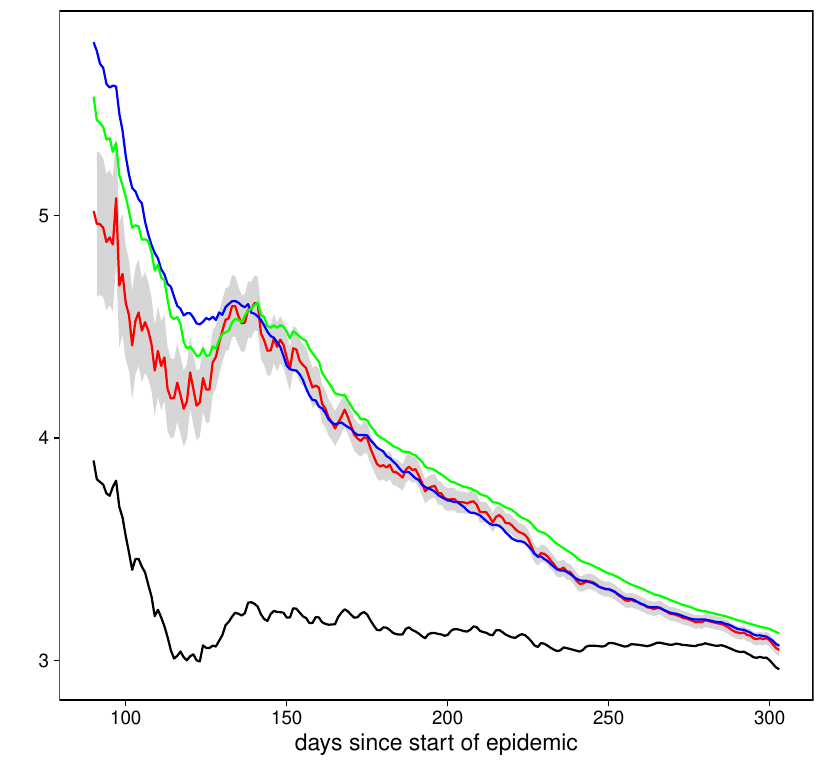}
		\caption{Estimated case fatality rate of COVID-19 in Argentina in 2020 computed by different methods. Shaded area corresponds to the union of the C.I. of $CFR(t)$ for each $t$. Black curve corresponds to  $CFR_N(t)$, red curve to $CFR(t)$, green curve to $CFR_G(t)$ and blue curve to $CFR_F(t)$.}		\label{fig:datosrealesarg}
	\end{figure}

		Figure \ref{fig:datosrealesarg} displays the estimated curves, together with the observed 
	$CFR_F(t)$. First, observe that both  $CFR_G(t)$  and  $CFR(t)$ do a much better job than the usual  $CFR_N(t)$ at estimating $cfr(t)$, since they are both much closer to $CFR_F(t)$. Second, we remark that $CFR_G(t)$ is nearer to $CFR_F(t)$  until around day $110$ but soon after that day, the biased nature of $CFR_G(t)$ becomes apparent. Other simulations (unreported here) show that  $CFR_G(t)$ has less variability than $CFR(t)$. However, since for both estimators the variance converges to zero as the number of observations goes to infinity, the weight of the variance in the bias-variance trade off gets smaller and smaller as the number of cases rises. By approximately day $150$, it becomes evident that $CFR_G(t)$ exhibits bias and converges to a value different from the intended estimation of $cfr(t)$.
	{Finally, it is noteworthy that the most realistic approach would have been to utilize the database reported on day $t$ for computing the estimators at each $t$ in the analysis, instead of employing the April 4th database for all estimations. Unfortunately, this was not feasible due to the unavailability of all databases from June to December 2020.
		However, we did have access to some databases from June 2020 and observed (in estimations not reported here) that the estimators did not perform as well as those reported in this study. The issue arose from delays in entering data, i.e., many individuals who were confirmed and deceased by day $t$ only appeared in the database for day $t$ as confirmed cases but not as deceased cases. Conversely, in a subsequent database (for instance, one from two months later), these individuals appeared correctly as diagnosed and deceased by day $t$.
			Addressing this data entry delay problem may involve the application of nowcasting techniques, as discussed in \cite{bastos2019modelling}. 
		
		We explored different estimators of $F$, assuming parametric models for this distribution, including regression of $Y$ on $X$, where $Y$ represents  the number of days from diagnosis to death, and $X$ represents the number of days from the beginning of the epidemic (March 3rd in this case) until the diagnosis date. We employed a linear model and different generalized linear models. However, for the majority of values of $t$, the regressions were non-significant. Additionally, it was necessary to use the fitted models to estimate $F_d$ for values of $d$ that were significantly beyond the range of the training data, leading to substantial extrapolation errors. We believe this issue can be addressed by employing techniques designed for censored data. However, this does not appear straightforward for this type of data and may be explored in future work.}



\section{Concluding remarks}\label{sec:conclu}

{Given the diagnosed cases of an epidemic disease, a statistical model is established for the outcomes of the disease for patients diagnosed on different days. Based on the model, the case fatality rate of the disease by day $t$ is defined as the probability of dying from the disease for a randomly chosen person among those diagnosed by day $t$, and an estimator for this rate is proposed. This estimator is based on the distribution of the times between confirmation and death of confirmed cases that die because of the disease..   

It is demonstrated that the proposed estimator is unbiased, consistent, and asymptotically normal, and asymptotic confidence intervals are derived. Both the estimator and confidence intervals for the case fatality rate can be computed during the course of the epidemic.

The excellent performance of the proposed estimator and confidence intervals for large samples is demonstrated in comparison to the estimator proposed by \cite{garske2009assessing} and the ``naive'' estimator reported daily during the COVID-19 epidemic. This is achieved through a Monte Carlo study and an analysis of the COVID-19 epidemic in Argentina during 2020.
The mean coverage of the asymptotic confidence intervals is computed as a function of the cumulative number of cases, and it is shown to closely approximate the nominal confidence level when the number of cases is large.

An estimator of the daily case fatality rate and an extension of the estimator proposed by \cite{garske2009assessing} that allows the distribution of the times between confirmation and death to change over time are also proposed.

To conclude, some limitations of our proposal and potential future directions of our work are mentioned. First, the computation of our proposed estimator requires more information than the simpler estimator proposed by \cite{garske2009assessing}, namely the number of confirmations and deaths each day. Second, the delay in entering data, usually present in epidemic outbreaks, may introduce extra bias and variability. This problem may be dealt with using nowcasting techniques, as demonstrated in \cite{bastos2019modelling}. This may be the subject of further work. Third, the difficulty of estimating $F$, the distribution of the time between confirmation and death.
In the real data analysis, different estimators of $F$ were studied, assuming parametric models for this distribution, including regression of $Y$ on $X$, where $Y$ represents the number of days from diagnosis to death, and $X$ represents the number of days from the beginning of the epidemic until the diagnosis date. A linear model and different generalized linear models were fitted. However, for the vast majority of values of $t$, the regressions were nonsignificant. Moreover, it was necessary to use the fitted models to estimate $F_d$ for values of $d$ that were significantly beyond the range of the training data, resulting in large extrapolation errors. We believe this problem can be addressed by employing techniques designed for censored data. However, this does not seem simple for this kind of data and may be the subject of further work.

	
 \renewcommand{\theequation}{A-\arabic{equation}}
 \renewcommand{\thetheorem}{A-\arabic{equation}}
 \renewcommand{\thesubsection}{A-\arabic{equation}}
\addcontentsline{toc}{section}{Appendices}
\renewcommand{\thesubsection}{\Alph{subsection}}

\section*{Appendix}

\subsection{Proofs}\label{apendice}

In this section, the consistency and asymptotic normality of $CFR(t)$ are proven. Consistency of $CFR(t)$ is defined as the convergence in probability of \linebreak 
\noindent $CFR(t)-cfr(t)$ to $0$, as $t\to \infty$.
Firstly, the Central Limit Theorem for triangular arrays is recalled. 
\begin{theorem*}\label{triangTCL}
	Suppose that for each $t\in \natu_0$, $X_{t,1},X_{t,2},\ldots, X_{t,r_t}$ are independent random variables. Let $S_t=X_{t,1}+X_{t,2}+\ldots+ X_{t,r_t}$. Suppose that $\esp(X_{t,k})=0$ for all $t$ and $k$, and that the variances $\esp(X_{t,k}^2)$ are finite. Call $\sigma_t^2=\mathbb{V}(S_t)$. If the Lindeberg condition is satisfied, \textsl{i.e.}:
	
	$$  \lim_{t \to \infty} \frac{1}{\sigma_t^2}\sum_{k = 1}^{r_t} \esp \left(X_{t,k}^2 \cdot \mathbf{1}_{\{ | X_{t,k} | > \varepsilon s_{t} \}} \right)= 0$$ for all $\varepsilon > 0$, where  $\mathbf{1}_{\{\cdots\}}$ is the indicator function, then the central limit theorem holds, i.e. $$\frac{S_t}{\sigma_t} \convdist \itN(0,1).$$
\end{theorem*}
\noindent For a proof of this theorem,  see Theorem 27.2 in \cite{billingsley2008probability}.

%
%
%
%
%
%
\noindent \textbf{Proof of Theorem \ref{teo}.} Even though weak consistency is a consequence of asymptotic normality, the proof of part $(i)$ is written because it is interesting in itself and has a simple proof independent of $(ii)$.

The estimator $ CFR(t)$ proposed for the case fatality rate, $cfr(t)$, is a sample average of the variables $$Z_{d,i}(t)= \frac{D_{d,i}(t)}{F_d(t-d)}.$$ Easy calculations show that $\esp(Z_{d,i}(t))=p_d$ and $$\mathbb{V}(Z_{d,i}(t))=\frac{p_d(1-p_dF_d(t-d))}{F_d(t-d)}$$ are finite. 

\noindent $(i)$ Since \textbf{A1} is satisfied, it is obtained that
$$\mathbb{V}(Z_{d,i}(t))\le \frac{1}{\inf_{d'}{F_{d'}(0)}}-\inf_{d'}p_{d'}\le \frac{1}{D},$$
for all $d$ and for all $i$. Then
$$\mathbb{V}(CFR(t))=\frac{\sum_{d=0}^{t}c_d \mathbb{V}(Z_{d,i}(t))}{r_t^2} \le  \frac{1}{D r_t}\stackrel{t\to\infty}{\longrightarrow} 0,$$
since $r_t\stackrel{t\to \infty}{\longrightarrow}\infty$. Since $CFR(t)$ is an unbiased estimator of $cfr(t)$, this implies that $CFR(t)-cfr(t) \convprob  0$.

\noindent $(ii)$ 
For each day $t$, the centered variables $Z_{d,i}(t)-p_d$ for $0\le d\le t$ and $1\le i\le c_d$, yield $r_t$ independent random variables of mean $0$. If these variables are renamed as $X_{t,1},\ldots,X_{t,r_t}$, it can be observed that the Lindeberg condition is satisfied. Subsequently, if $S_t=X_{t,1}+X_{t,2}+\ldots+ X_{t,r_t}$ and $\sigma_t^2=\mathbb{V}(S_t)$, the convergence in distribution 
$$\frac{S_t}{\sigma_t}\convdist \itN(0,1)$$ 
is achieved. The proof concludes by noting, through elementary calculations, that 
$$
\frac{CFR(t)-cfr(t)}{\sqrt{\mathbb{V}(CFR(t))}}=\frac{S_t}{\sigma_t}\,.
$$

Let us see that  the variables $\{Z_{d,i}(t)-p_d\}_{d,i}$ satisfy the Lindeberg condition.  It is observed that 
$$\mathbb{V}(Z_{d,i}(t))\ge (\inf_{d'} p_{d'})(1-\sup_{d'} p_{d'})=I(1-S),$$
for all $d$ and for all $i$. This implies $$\sigma_t^2=\sum_{d=0}^t\sum_{i=1}^{c_d}\mathbb{V}(Z_{d,i}(t))\ge r_t I(1-S)$$ 
and then $\sigma_t\to \infty$ as $r_t\to\infty$. Since \textbf{A2} and \textbf{A3} hold, the lower bound of $\sigma_t^2$ is positive. It is observed that the variable $Z_{d,i}(t)-p_d$ takes only two possible values: $-p_d$ and $1/F_d(t-d)-p_d$. so it is obtained that $$|Z_{d,i}(t)-p_d)|\le \max \left\{1,\frac{1}{\displaystyle\inf_d{F_d(0)}}-\inf_{d'}p_{d'}\right\}\le \max \left\{1,\frac 1D\right\}=M.$$ Given $\varepsilon>0$, if $r_t$ is large enough, then $\varepsilon\cdot \sigma_t>M$ and 
${\mathbf{1}_{\{ | Z_{d,i}(t)-p_{d}| > \varepsilon \sigma_t \}}=0}$ for all $d$ and $i$, and then the Lindeberg condition
$$  \lim_{t \to \infty} \frac{1}{\sigma_t^2}\sum_{d=0}^t\sum_{i=1}^{c_d} \esp
 \left((Z_{d,i}(t) - p_d)^2 \cdot \mathbf{1}_{\{ | Z_{d,i}(t) - p_{d} | > \varepsilon \sigma_t \}} \right)= 0$$ is satisfied.	




\section{Funding and  Conflicts of interests}
This work was partially supported by Grants PICT 2018-00740 and PICT-201-0377 from  Agencia Nacional de Promoción Científica y Tecnológica at Buenos Aires, Argentina and Grant 20020170100330BA from Universidad de Buenos Aires.

The authors declare that they have no conflicts of interest.
	\bibliography{main-2024-01-23}

	\end{document}